\documentclass[a4paper,11pt]{article}
\pdfoutput=1
\usepackage{jcappub}
\usepackage[T1]{fontenc}
\usepackage{bm}
\usepackage{todonotes}

\title{\boldmath Neutrino physics with the PTOLEMY project: active neutrino properties and the light sterile case}
\author[a,b]{M.G.~Betti,}
\author[c]{M.~Biasotti,}
\author[d]{A.~Bosc{\'a},}
\author[d]{F.~Calle,}
\author[e]{N.~Canci,}
\author[a,b]{G.~Cavoto,}
\author[f,g]{C.~Chang,}
\author[h]{A.G.~Cocco,}
\author[i,j]{A.P.~Colijn,}
\author[k]{J.~Conrad,}
\author[e]{N.~D'Ambrosio,}
\author[i,l]{N.~De~Groot,}
\author[m]{P.F.~de~Salas,}
\author[n]{M.~Faverzani,}
\author[k]{A.~Ferella,}
\author[n]{E.~Ferri,}
\author[o]{P.~Garcia-Abia,}
\author[p]{I.~Garc{\'i}a-Cort{\'e}s,}
\author[q]{G.~Garcia~Gomez-Tejedor,}
\author[r]{S.~Gariazzo,}
\author[c]{F.~Gatti,}
\author[s]{C.~Gentile,}
\author[n]{A.~Giachero,}
\author[k]{J.E.~Gudmundsson,}
\author[t]{Y.~Hochberg,}
\author[u]{Y.~Kahn,}
\author[v]{A.~Kievsky,}
\author[w]{M.~Lisanti,}
\author[a,b]{C.~Mancini-Terracciano,}
\author[h]{G.~Mangano,}
\author[v,x]{L.E.~Marcucci,}
\author[a,b]{C.~Mariani,}
\author[d]{J.~Mart{\'i}nez,}
\author[e,y]{M.~Messina,}
\author[o]{A.~Molinero-Vela,}
\author[z]{E.~Monticone,}
\author[p]{A.~Moro{\~n}o,}
\author[n]{A.~Nucciotti,}
\author[b]{F.~Pandolfi,}
\author[e]{S.~Parlati,}
\author[r]{S.~Pastor,}
\author[d]{J.~Pedr{\'o}s,}
\author[aa]{C.~P{\'e}rez~de~los~Heros,}
\author[h,ab]{O.~Pisanti,}
\author[a,b]{A.D.~Polosa,}
\author[n]{A.~Puiu,}
\author[a,b]{I.~Rago,}
\author[s]{Y.~Raitses,}
\author[z]{M.~Rajteri,}
\author[e]{N.~Rossi,}
\author[o]{I.~Rucandio,}
\author[o]{R.~Santorelli,}
\author[y]{K.~Schaeffner,}
\author[w]{C.G.~Tully,}
\author[v]{M.~Viviani,}
\author[w]{F.~Zhao,}
\author[f,ac]{K.M.~Zurek}

\affiliation[a]{Universit{\`a} degli Studi di Roma La Sapienza, Roma, Italy}
\affiliation[b]{INFN Sezione di Roma, Roma, Italy}
\affiliation[c]{Universit{\`a} degli Studi di Genova e INFN Sezione di Genova, Genova, Italy}
\affiliation[d]{Universidad Polit{\'e}cnica de Madrid, Madrid, Spain}
\affiliation[e]{INFN Laboratori Nazionali del Gran Sasso, L'Aquila, Italy}
\affiliation[f]{Department of Physics, University of California, Berkeley, CA, USA}
\affiliation[g]{Argonne National Laboratory, Chicago, IL, USA}
\affiliation[h]{INFN Sezione di Napoli, Napoli, Italy}
\affiliation[i]{Nationaal instituut voor subatomaire fysica (NIKHEF), Amsterdam, Netherlands}
\affiliation[j]{University of Amsterdam, Amsterdam, Netherlands}
\affiliation[k]{Stockholm University, Stockholm, Sweden}
\affiliation[l]{Radboud University Nijmegen, Nijmegen, Netherlands}
\affiliation[m]{The Oskar Klein Centre for Cosmoparticle Physics, Department of Physics, Stockholm University, SE-10691 Stockholm, Sweden}
\affiliation[n]{Universit{\`a} degli Studi di Milano-Bicocca e INFN Sezione di Milano-Bicocca, Milano, Italy}
\affiliation[o]{Centro de Investigaciones Energ{\'e}ticas, Medioambientales y Tecnol{\'o}gicas (CIEMAT), Madrid, Spain}
\affiliation[p]{Laboratorio Nacional de Fusi{\'o}n, CIEMAT, Madrid, Spain}
\affiliation[q]{Instituto de F{\'\i}sica Fundamental, Consejo Superior de Investigaciones Cientificas (CSIC), Madrid, Spain}
\affiliation[r]{Instituto de F{\'\i}sica Corpuscular  (CSIC-Univ. de Val{\`e}ncia), Valencia, Spain}
\affiliation[s]{Princeton Plasma Physics Laboratory, Princeton, NJ, USA}
\affiliation[t]{Racah Institute of Physics, Hebrew University of Jerusalem, Jerusalem, Israel}
\affiliation[u]{Kavli Institute of Cosmological Physics, University of Chicago, Chicago, IL, USA}
\affiliation[v]{INFN Sezione di Pisa, Pisa, Italy}
\affiliation[w]{Department of Physics, Princeton University, Princeton, NJ, USA}
\affiliation[x]{Universit{\`a} degli Studi di Pisa, Pisa, Italy}
\affiliation[y]{Gran Sasso Science Institute (GSSI), L'Aquila, Italy}
\affiliation[z]{Istituto Nazionale di Ricerca Metrologica (INRiM), Torino, Italy}
\affiliation[aa]{Uppsala University, Uppsala, Sweden}
\affiliation[ab]{Universit{\`a} degli Studi di Napoli Federico II, Napoli, Italy}
\affiliation[ac]{Lawrence Berkeley National Laboratory, University of California, Berkeley, CA, USA}

\abstract{
The PTOLEMY project aims to develop a scalable design for a Cosmic
Neutrino Background (CNB) detector, the first of its kind and the only
one conceived that can look directly at the image of the Universe encoded in neutrino
background produced in the first second after the Big Bang.
The scope of the
work for the next three years is to complete the conceptual design of this detector
and to validate with direct measurements that the
non-neutrino backgrounds are below the expected cosmological signal.
In this paper we discuss in details the theoretical aspects of the
experiment and its physics goals.
In particular, we mainly address three issues.
First we discuss the sensitivity of PTOLEMY to the standard neutrino
mass scale.
We then study the perspectives of the experiment to detect the CNB via neutrino
capture on tritium as a function of the neutrino mass scale and the energy
resolution of the apparatus.
Finally, we consider an extra sterile neutrino with mass in the eV range, coupled to the active states via oscillations, which has been advocated in view of neutrino oscillation anomalies.
This extra state would contribute to the tritium decay spectrum,
and its properties, mass and mixing angle, could be studied by analyzing
the features in the beta decay electron spectrum.
}

\newcommand{\mHe}{m_{^3{\rm He}}}
\newcommand{\mH}{m_{^3{\rm H}}}
\newcommand{\Eendz}{E_{\rm end,0}}

\newcommand{\e}[1]{\ensuremath{\times10^{#1}}}
\newcommand{\mli}{\ensuremath{m_{\rm lightest}}}
\newcommand{\ssq}[1]{\ensuremath{\sin^2\theta_{#1}}}
\newcommand{\dmsq}[1]{\ensuremath{\Delta m^2_{#1}}}

\newcommand{\ordfit}[2]{\ensuremath{\widehat{\rm #1}/{\rm #2}}}
\newcommand{\lnB}[1]{\ensuremath{\ln\mathcal{B}_{#1}}}
\newcommand{\bev}[1]{\ensuremath{\mathcal{Z}_{#1}}}

\begin{document}
\maketitle

\section{Introduction}

The Universe has expanded by a factor of over one billion between the early thermal epoch known as the neutrino decoupling stage and the present day.
We observe this dynamics in many forms: the recession of galaxies (Hubble expansion), the dim afterglow of the hot plasma epoch, the Cosmic Microwave Background (CMB), and the abundances of light elements during Big Bang Nucleosynthesis (BBN).
The epoch of neutrino decoupling produced another pillar of confirmation, the Cosmic Neutrino Background (CNB), perhaps one of the most important not yet {\it directly} probed predictions of the standard cosmological model.
Because of the similarities shared with the CMB, its properties are theoretically expected to be very close to those of the photon background.
In the so-called instantaneous decoupling limit, i.e.\ assuming that neutrino weak interactions become slower than the Hubble rate {\it instantaneously}, we have a simple expression for the average number density per neutrino state today
\begin{equation}
n_0
=
\frac{3 \zeta (3)}{4\pi^2} T_{\nu,0}^{3} = 56\,\mathrm{cm}^{-3}\,,
\label{n0}
\end{equation}
with a neutrino temperature $T_{\nu,0} \simeq 1.95\,\mathrm{K}$, and where the subscript '0' refers to the present epoch.
Actually, the CNB spectrum is expected to deviate from the one obtained considering a perfect Fermi-Dirac distribution at the percent level \cite{Dolgov:1997mb,Mangano:2005cc,deSalas:2016ztq}.
The reason is the partial overlap of the last stage of neutrino decoupling and the first instants of $e^\pm$ annihilations in the primeval plasma.
The same density is also shared by antineutrino species, assuming a vanishing neutrino-antineutrino asymmetry.
This assumption is well justified since the ratio of neutrino asymmetry is strongly constrained by comparing the theoretical description of the BBN stage with astrophysical determinations of both primordial $^4$He and deuterium nuclei, $-0.71 \leq L_\nu \leq 0.054$ \cite{Mangano:2011ip,Castorina:2012md},
where $L_\nu$ is the total neutrino asymmetry, summed over the three flavors, normalized by the photon number density $n_\gamma$.
Actually, the neutrino asymmetry in the standard baryogenesis through leptogenesis model is expected to be much smaller than this bound, of the order of the baryon density today,
$n_B/n_\gamma \sim 6 \cdot 10^{-10}$.

The CNB acts as a source of gravity through its pressure and energy density, which are usually cast in terms of the effective number of relativistic neutrinos
\begin{equation}
N_{\rm eff}
=
\frac87 \left( \frac{11}{4} \right)^{4/3} \frac{\rho_\nu}{\rho_\gamma}
\label{neff}\,,
\end{equation}
where $\rho_\nu$ ($\rho_\gamma$) is the neutrino (photon) energy density.
The theoretical value for this parameter including the small non thermal distortion to the CNB is $N_{\rm eff}= 3.045$~\cite{Mangano:2005cc,deSalas:2016ztq},
in good agreement with the bounds coming from BBN and Planck observations \cite{Aghanim:2018eyx},
which thus constrain any extra contribution to the radiation energy density due to exotic relativistic species and/or non standard features in the neutrino momentum distribution.

We know from neutrino flavor oscillation experiments, performed over more than thirty years
\cite{Fukuda:1998mi,Ahmad:2001an,Ahmad:2002jz},
that at least two of the three standard neutrinos are massive particles.
While the (squared) mass differences are very well measured today
\cite{deSalas:2017kay,Capozzi:2018ubv,Esteban:2018azc},
we are still ignorant of the absolute mass scale, set by the lightest neutrino mass eigenstate, and the mass ordering, which is related to a possible hierarchy of the mass spectrum (see e.g.\ \cite{Capozzi:2017ipn,deSalas:2018bym,Esteban:2018azc}).
One of the best ways to measure the absolute neutrino mass scale is through the study of the endpoint of the $\beta$ decay or the electron capture decay of some convenient atoms, such as tritium and holmium.
The bound on the effective mass entering the tritium $\beta$ decay, $m_\beta$, obtained by studying the emitted electron energy spectrum near the endpoint, is presently $m_\beta < 2$ eV \cite{Aseev:2011dq,Kraus:2012he,Tanabashi:2018oca}, but the KATRIN experiment,
currently taking data, is expected to improve this bound by an order of magnitude in the near future, or obtain a measurement if $m_\beta$ is at the level of fractions of eV \cite{Angrik:2005ep,Sibille:2019qsn}.
Other future experiments such as ECHo~\cite{Eliseev:2015pda}, HOLMES~\cite{Nucciotti:2018vyc} or Project-8~\cite{Esfahani:2017dmu} will also study the electron capture or $\beta$ decay endpoint of holmium or atomic tritium, respectively, to obtain constraints on the neutrino mass scale.
Interestingly, this range of neutrino masses can also be scrutinized using cosmological observations.
We will present a short summary of this issue in the next Section, as well as the present status of oscillations, neutrino mass bounds and mass ordering. 
Moreover, since massive neutrinos became non-relativistic at a certain time, depending on their absolute masses, they can be gravitationally trapped under the effect of large enough gravitational potentials \cite{Ringwald:2004np, deSalas:2017wtt,Zhang:2017ljh}, and this enhances the local density at the Earth with respect to the homogeneous value $n_0$.
This is quite important for their direct detection perspectives, since the signal rate is proportional to their density.

Experimental advances both in the understanding of massive neutrino physics and in techniques of high sensitivity instrumentation have opened up new opportunities to directly detect the CNB, an achievement which would profoundly confront and extend the sensitivity of precision cosmology data.
The aim of the PTOLEMY project \cite{Baracchini:2018wwj} is to develop a scalable design for a CNB detector, the first of its kind and the only one conceived that can look directly at the neutrino background.
The scope of the work for the next three years is to complete the blueprint of the Cosmic Neutrino detector and to validate that the non-neutrino backgrounds are below the expected signal from the Big Bang with a direct measurement.
An array of detectors of this design could reach discovery sensitivity for the CNB.
The number and deployment of these detectors around the world will depend on the next phase of PTOLEMY developments, described in \cite{Baracchini:2018wwj}.
Yet, the physics case of the experiment is quite wide, including, as another major goal, the measurement of the standard neutrino absolute mass  scale, in a way similar to that of the KATRIN experiment \cite{Angrik:2005ep}.
Moreover, some non standard scenarios could be tested, also in the preliminary phases of development of PTOLEMY.
For example, an interesting issue concerns the physics of sterile neutrino states with masses in the eV range.
Sterile eV neutrinos which mix with active states have been suggested since the LSND results \cite{Aguilar:2001ty} to solve some anomalies in neutrino oscillation experiments (short baseline data, reactor anomaly and gallium anomaly) \cite{Abazajian:2012ys,Gariazzo:2015rra,Giunti:2019aiy} and their existence would have cosmological implications, see e.g.\ \cite{Mirizzi:2013gnd,Boser:2019rta}.

In this paper we discuss in detail the theoretical aspects of the experiment and its neutrino physics goals.
We study the sensitivity of PTOLEMY on neutrino mass detection, CNB detection and signals or bounds on sterile neutrinos with masses in the eV range, as a function of the expected energy resolution on the outgoing electrons and the employed mass of the tritium source.
The scenario of keV neutrinos as warm dark matter candidates and their imprint on tritium spectrum,
which appears well below the $Q$ value of the decay, will be considered elsewhere.

The paper is structured as follows.
In Section~\ref{sec:theory} we review the to date information on neutrino masses, both from tritium decay and cosmological observables, mass ordering and oscillations,
and on possible exotic neutrino states or properties, such as the possible existence of a light sterile neutrino.
In Section~\ref{sec:beta} we introduce the formalism used in the analysis and describe how the finite electron energy resolution is taken into account.
In Section~\ref{sec:analysis} we illustrate our method, the Bayesian analysis we use to obtain forecasts of the experiment sensitivity.
In Section~\ref{sec:mass} we then apply this analysis to the lightest neutrino mass parameter, showing the PTOLEMY discovery potential for the neutrino mass scale as a function of the tritium sample mass.
We then study in Section~\ref{sec:cnb} the perspectives of the experiment to detect the CNB via neutrino
capture on tritium as a function of the neutrino mass scale and the energy resolution of the apparatus.
The scenario of an extra sterile neutrino with mass in the eV range, coupled to the active states via oscillations, is considered in Section~\ref{sec:sterile}, where we describe how its properties, mass and mixing angle, could be constrained by analyzing the $\beta$ decay electron spectrum.
Finally, we present our conclusions and outlooks in Section \ref{sec:conclusions}.

\section{Theoretical context}
\label{sec:theory}

As already introduced, the cleanest determination of the absolute scale of neutrino masses
would proceed from a precise observation of the electron or positron spectrum close to the end point
of $\beta$ decay.
Current best limits on the effective electron \emph{antineutrino} mass
come from the observations of tritium decay in the Troitsk \cite{Aseev:2011dq} and Mainz \cite{Kraus:2012he} experiments.
As discussed in the next section, each neutrino%
\footnote{We assume neutrinos and antineutrinos share the same mass,
although there is no experimental confirmation of this fact.}
mass eigenstate contributes to the suppression of the electron spectrum,
but when the energy resolution is not sufficient to discriminate the different contributions it is safe
to parameterize the effective neutrino $\beta$ decay mass as:
\begin{equation}
m_\beta^2
=
\sum_i |U_{ei}|^2 m_i^2
\,,
\end{equation}
which is the experimentally determined quantity.
It depends on the mixing matrix elements that describe the fraction of electron flavor for each mass eigenstate, $U_{ei}$,
and on the mass of the $i$-th neutrino eigenstate, $m_i$.
When the $m_i\simeq m_\nu$ are very similar and the neutrino masses are degenerate,
the above definition becomes $m_\beta\simeq m_\nu$.
The current best limit is $m_\beta<2$~eV, at 95\% CL \cite{Aseev:2011dq,Kraus:2012he,Tanabashi:2018oca}.
In the incoming years, several experiments are expected to provide new and more stringent bounds.
KATRIN \cite{Angrik:2005ep,Sibille:2019qsn} started taking data in 2018 and the first results are expected soon,
while the final sensitivity with 5~years of data will allow to reach $m_\beta\lesssim0.2$~eV at 90\% CL.
Other experiments under development include
Project-8~\cite{Esfahani:2017dmu}, which is expected to reach the sensitivity level of $m_\beta\lesssim40$~meV at 90\% CL using atomic tritium,
and the two holmium experiments ECHo~\cite{Eliseev:2015pda} and HOLMES~\cite{Nucciotti:2018vyc},
which aim at detecting the \emph{neutrino} mass using the electron capture decay of $^{163}$Ho
with a sensitivity around the eV.

Another, indirect, probe of neutrino masses comes from cosmological observables.
In fact, neutrinos have cooled during the expansion of the Universe and their presence and masses can be indirectly felt through the action of their diminishing thermal velocities on large-scale structure formation.
In particular, Planck data constrain the sum of the three neutrino masses mainly via the lensing power spectrum and the lensing effects on CMB anisotropies.
The bound on the sum of the neutrino masses is presently in the range $\sum_i m_i < (0.24-0.54)$~eV (95\% CL) when considering CMB observations only \cite{Aghanim:2018eyx},
depending on the CMB data which are used in the analysis (see also \cite{Ade:2015xua,Lattanzi:2017ubx}).
CMB probes are not the ideal way to constrain the neutrino masses,
as neutrinos were relativistic at the time of photon decoupling and the effect of their masses
is mainly imprinted through the late-time evolution of the CMB spectrum.
For this reason, combinations of CMB data with low redshift probes such as determinations of the Baryon Acoustic Oscillations (BAO),
of the matter power spectrum at late times
or of other probes such as from the absortion spectrum measured from Lyman-$\alpha$ forests
provide a stronger constraint on the sum of the neutrino masses.
When CMB observations are combined with BAO, for example, the limits tighten thanks to the combination of observations that are relevant at different epochs,
and the limits become $\sum_i m_i < (0.12-0.16)$~eV (95\% CL) \cite{Aghanim:2018eyx}.
From Lyman-$\alpha$ data alone, on the other hand, a bound of $\sum_i m_i < 0.8$~eV (95\% CL) is found \cite{Yeche:2017upn},
while Lyman-$\alpha$ in combination with CMB observations leads to a limit on the sum of neutrino masses of $\sum_i m_i < (0.12- 0.14)$~eV (95\% CL) \cite{Palanque-Delabrouille:2015pga,Yeche:2017upn}.
Future observations of the matter power spectrum, such as those inferred by galaxy surveys like Euclid \cite{Amendola:2016saw},
will allow to obtain a determination of $\sum_i m_i$ to a precision of $\sim0.02$~eV in combination with Planck \cite{Brinckmann:2018owf}.
One must always remember that cosmological constraints on the absolute neutrino mass are obtained indirectly
under the assumption of a specific cosmological model.
The results quoted above are for example derived assuming the simplest case,
where the Universe evolution is described by the six parameters of the $\Lambda$CDM model
plus the sum of the neutrino masses.
When more parameters are varied, the limits can be relaxed up to a factor three
(see e.g.~\cite{Ade:2015xua,DiValentino:2015ola,Gariazzo:2018meg}).
Bayesian techniques allow to marginalize over the possible extensions of the minimal model,
and a robust limit can be obtained relaxing the Planck ones in the $\Lambda$CDM+$\sum_i m_i$ model by approximately 50\% \cite{Gariazzo:2018meg}.

Another unknown related to neutrino masses is their ordering,
which can be normal if the lightest neutrino is the one with the largest mixing with the electron flavor,
or inverted in the opposite case.
The mass ordering is defined by the undetermined sign of the mass splitting that mostly affects atmospheric
neutrino oscillations, $\Delta m^2_{31}$.
The sign of such mass difference can be only determined by oscillation experiments which are sensitive to matter effects, such as atmospheric or long-baseline accelerator neutrino experiments.
The neutrino mass ordering is very important to interpret the number of events which PTOLEMY can observe,
as we will discuss in the next section.
At present, the strongest preference for one of the two possible orderings by a single experiment
comes from Super-Kamiokande, which favors normal ordering at $\sim 1.4\sigma$ \cite{Jiang:2019xwn}.
Results from long-baseline oscillation experiments such as
NO$\nu$A \cite{Acero:2019ksn}
or T2K \cite{Abe:2018wpn},
instead, provide a preference for normal ordering
around the 2$\sigma$ level (each) only when a prior on $\theta_{13}$ from reactor experiments is adopted.
When all relevant neutrino oscillation experiments are combined in a global fit, the preference rises up to $\sim 3\sigma$ \cite{Capozzi:2017ipn,deSalas:2017kay,Esteban:2018azc}.
The addition of data from other probes (such as $0\nu\beta\beta$, CMB, BAO or a prior on $H_0$) only improves slightly the preference towards a normal ordering \cite{Gariazzo:2018pei,deSalas:2018bym}.
A definite determination of the mass ordering is expected in the next few years when the currently ongoing experiments
will improve the statistics, although is unlikely that any of them will be able to reject the wrong ordering alone.
In the near future, however, the ORCA experiment from the KM3NeT collaboration will improve significantly the sensitivity to the neutrino mass ordering.
ORCA, which is expected to deploy a working set of strings in the next years,
can reach a $5\sigma$ preference in three years provided that the true ordering is normal \cite{Adrian-Martinez:2016fdl}.
In the case of an inverted ordering, the reached sensitivity in three years will be of $\sim 3\sigma$,
which will increase in combination with other experiments.
The final mass ordering determination will be performed by DUNE, expected to start in 2026,
which will reach the $5\sigma$ significance regardless of the true ordering after 7 years of data taking \cite{Acciarri:2015uup}.

As we will detail in the next Section, the rate of CNB capture on a tritium nucleus,
the process which PTOLEMY will exploit, can be written as
\begin{equation}\label{eq:nucapture_events_i-0}
\Gamma_{\rm{CNB}}
= \sum^{N_\nu}_{i=1}
N_T \,
|U_{ei}|^2\,
\bar\sigma\, v_\nu
f_{c,i}\,
n_{0}\,,
\end{equation}
where the sum is over neutrino mass eigenstates,
$n_{0}$ is the average neutrino number density on large scales, see Eq.~\eqref{n0},
$f_{c,i} \geq 1$ are the clustering factors,
defined as the ratio between the local and the average number density $n_0$,
that code the local overdensity of these particles due to the gravitational attraction of our galaxy,
and $v_\nu$ the neutrino velocity in the Earth frame.
Finally, the quantity $\bar{\sigma}$ is the average cross section for neutrino capture.
As discussed in the following, the neutrino mass ordering enters the above equation through
the various $U_{ei}$, which are different when considering a normal or inverted mass ordering.

Apart from the obvious dependence of the shape of emitted electron energy on neutrino masses,
the latter also affects the $f_{c,i}$ parameters, which monotonically increase with $m_i$.
Since at least two of the neutrino mass eigenstates are non relativistic in the recent times of structure formation, indeed,
relic neutrinos tend to cluster in overdensity regions, such as the Milky Way or the Virgo cluster, to which our galaxy belongs.
As shown in \cite{Ringwald:2004np,deSalas:2017wtt,Zhang:2017ljh} this leads to an increased capture rate at Earth,
which can be larger by a factor 10-20\% for neutrino masses around 60~meV
or up to 200\% for masses of $150$~meV, also depending on the assumed matter profile of the halo where the clustering occurs.
Focusing only on one single case for the Milky Way composition,
one can write the clustering factor as a function of the neutrino mass using a power-law.
Considering a generalized Navarro-Frenk-White \cite{Navarro:1995iw} profile
and baryon content of the galaxy as in \cite{deSalas:2017wtt},
one can find \cite{Zhang:2017ljh}:
\begin{equation}
\label{eq:clust_fac_mi}
f_{c,i}
=
76.5\left(\frac{m_i}{{\rm eV}}\right)^{2.21}
\,.
\end{equation}

The nature of neutrinos is still unknown,
i.e.\ we ignore if they are Dirac or Majorana (self-conjugated) particles.
As well known, neutrinoless double $\beta$ decay is the most promising way to experimentally answer this issue.
Interestingly, the expected event rate of CNB capture on tritium is in general larger
for non-relativistic Majorana neutrinos with respect to the Dirac case~\cite{Long:2014zva,Roulet:2018fyh}.
The reason is that, when neutrinos become non-relativistic while free-streaming, helicity is conserved contrary to chirality.
In the Dirac case, this leads to a population of half the original amount of left-handed neutrinos that are left-chiral, and therefore able to be captured in tritium,
while in the Majorana case the original right-handed neutrinos also contribute with a developed left-chiral component,
which amounts to a twice larger local density of relic neutrinos that can be detected with respect to the Dirac case.
The total effect is not always a factor two in the event rate of Majorana neutrinos with respect to Dirac ones.
Depending on the mass hierarchy, if the lightest neutrino has an extremely small mass (below $\sim 1$ meV)
and is still relativistic today, its capture rate will be the same for both cases \cite{Roulet:2018fyh},
and the increase in the total event rate is much smaller than two.
A CNB detection with significant statistics would be a further way to understand the neutrino nature, though this goal seems quite demanding,
in particular because the effect is only visible in the event rate, which also depend, for example, on the clustering factors.

The event rate would be also modified by the presence of neutrino interactions beyond the standard weak processes, as those predicted in most of the extended theoretical models where neutrinos acquire mass. The characteristics of these non-standard interactions (NSI) involving neutrinos
depend on the specific model and the relevant operators \cite{Cirigliano:2013xha,Farzan:2017xzy}, but their existence could alter the event rate of relic cosmological neutrinos. For instance, in~\cite{Arteaga:2017zxg} it was shown that charged-current NSI involving Dirac neutrinos could change the capture rate on tritium in a PTOLEMY-like detector by a factor between 0.3 to 2.2, while in the Majorana case the possible variation would be restricted to a few percent.
Thus, also this kind of hypothetical interactions would affect the information that can be inferred on the neutrino nature.

Let us now consider a more exotic scenario with an extra neutrino mass eigenstate $\nu_4$, with mass around an eV,
mainly mixed with a new sterile neutrino flavor~\footnote{
Since the active neutrino flavors are expected to have a small mixing with the fourth mass eigenstate in order to preserve the phenomenology of the three active neutrino mixing,
the sterile neutrino flavor should mix almost only with the fourth mass eigenstate.
For this reason it is approximately correct to say that the sterile neutrino has a mass $m_s\simeq m_4\simeq1$~eV.
}.
The considered mass comes from the fact that a number of anomalies in neutrino oscillation experiments
could be solved by the presence of a light sterile neutrino with a mass around this scale.
The first anomaly was published by the LSND experiment \cite{Aguilar:2001ty},
the result of which was immediately criticized for being incompatible with neutrino oscillations \cite{Maltoni:2002xd}.
After the discovery of the gallium \cite{Abdurashitov:2009tn}%
\footnote{See also \cite{Kostensalo:2019vmv}.}
and reactor \cite{Mention:2011rk} anomalies,
more experimental efforts were planned to study more carefully the problem and obtain a final conclusion.
Nowadays, the situation is unclear due to a tension between the disappearance measurements,
including both the electron and muon neutrino channels,
and the appearance observations.
The tension arises from the fact that the reactor antineutrino experiments (discussed in details in the following paragraph)
seem to prefer a non-zero mixing between active and sterile neutrinos through the mixing matrix element $U_{e4}$,
but in the muon neutrino disappearance channel, which is probed mainly by
accelerator neutrino experiments as MINOS/MINOS+ \cite{MINOS:2016viw,Adamson:2017uda}
or atmospheric neutrino detectors such as
IceCube \cite{TheIceCube:2016oqi}
and DeepCore \cite{Aartsen:2017bap},
no oscillations have been observed.
As a consequence, we have only upper limits on the mixing matrix element $U_{\mu4}$.
On the other hand, neutrino appearance experiments as
LSND \cite{Aguilar:2001ty} and
MiniBooNE \cite{Aguilar-Arevalo:2018gpe},
which are sensitive to the product $|U_{e4}|^2|U_{\mu4}|^2$, observed an excess of neutrino events
that cannot be explained using the values of $U_{e4}$ and $U_{\mu4}$ inferred by disappearance probes.
The tension is known since many years (see e.g.~\cite{Gariazzo:2015rra,Giunti:2019aiy,Diaz:2019fwt,Boser:2019rta}),
but its significance has increase even more with the most recent MINOS+ and MiniBooNE results
\cite{Dentler:2018sju,Gariazzo:2017fdh}.

Since in this paper we are only interested in the effect that a sterile neutrino may have on the $\beta$ spectrum
or on neutrino capture events,
which can be described using only its mass and the mixing with electron flavor ($U_{e4}$),
we will only focus on the neutrino oscillation constraints that come from the electron (anti)neutrino disappearance channel,
which is sensitive to the squared mass difference $\dmsq{41}$ and the mixing matrix element $U_{e4}$.
One among the best approaches to distinguish the effect of new neutrino oscillations
from the presence of other systematic uncertainties,
like for example a wrong theoretical spectrum of reactor antineutrinos,
is to measure the flux at different distances from the source and to consider their ratios to do the analyses.
Since this method decouples the neutrino oscillation effects from the theoretical description of the initial flux,
it is referred to as ``model-independent approach''.
Nowadays, the strongest constraints come from the
NEOS \cite{Ko:2016owz},
DANSS \cite{Alekseev:2018efk},
Neutrino-4 \cite{Serebrov:2018vdw},
PROSPECT \cite{Ashenfelter:2018iov}
and STEREO \cite{Almazan:2018wln} experiments,
which all use a model-independent approach considering distances between 6 and 25~m from the reactor core.
Among these experiments, Neutrino-4 is the one that claimed to have observed active-sterile oscillations
with the highest significance \cite{Serebrov:2018vdw},
but the rather large values of the mixing parameters at the best-fit
are excluded at more than 95\% CL by PROSPECT \cite{Ashenfelter:2018iov}.
The Neutrino-4 collaboration did not discuss their compatibility with the best-fit found in \cite{Alekseev:2018efk},
for which the DANSS collaboration reports an improvement of the fit
with respect to the standard three neutrino oscillation paradigm with a $\Delta\chi^2\simeq13$.
This value corresponds to a $\simeq2.8\sigma$ preference for 3+1 neutrino oscillations
according to the preliminary results presented at the Neutrino 2018 conference \cite{Egorov:2018nu},
and is in excellent agreement with the NEOS results \cite{Ko:2016owz}
and the first PROSPECT results \cite{Ashenfelter:2018iov}.
When the DANSS and NEOS results, which are not in tension with any other known observation,
are analysed in a combined fit,
the preferred value for the new mass splitting and mixing with the electron flavor are
$\dmsq{41}\simeq1.29$~eV$^2$ and $|U_{e4}|^2=s^2_{14}\simeq0.012$~\cite{Gariazzo:2018mwd,Dentler:2018sju}.
Since more data are expected from the already mentioned experiments,
a final result on the existence of short-baseline oscillations at reactors is expected soon.

From the cosmological point of view, data from both BBN and CMB are incompatible with a fully thermalized sterile neutrino
with an eV mass and a mixing angle as required to solve the oscillation anomalies in the three neutrino picture
(see e.g.~\cite{Gariazzo:2019gyi,Boser:2019rta}).
In fact, this would turn into a larger radiation content in the early universe,
a faster expansion rate given by the Hubble factor,
which would change both the amount of primordial deuterium (and to a less extent of $^4$He) produced during BBN \cite{Saviano:2014esa},
the relative height of the first acoustic peak and the damping tail in the CMB power spectrum \cite{Tanabashi:2018oca,Mirizzi:2013gnd,Gariazzo:2015rra}.
In particular, a comparison of primordial deuterium observations with the theoretical analysis constrains the sterile state number density to a factor less than 0.8 (at 95\% C.L.) with respect to the active neutrino one, $n_0=56$~cm$^{-3}$.
We have obtained this result using the BBN public code \texttt{PArthENoPE 2.0} \cite{Consiglio:2017pot} and the most recent determination of deuterium, $^2$H/H $= (2.527 \pm 0.030) \cdot 10^{-5}$ \cite{Cooke:2017cwo},
exploiting the theoretical ab-initio calculation of the $d(p,\gamma)^3$He cross section from \cite{Marcucci:2015yla}.

In the recent years, models have been considered to relax this tension between a relic density of sterile neutrinos and cosmological obervables, using mechanisms which reduce the production of sterile states through oscillations.
One possibility is to consider asymmetries in the active neutrino sector.
In fact, while in a neutrino symmetric bath a thermal population of the sterile state would quickly grow, allowing for primordial neutrino asymmetries of order $L_\nu \geq {\cal O}(10^{-2})$ a self-suppression as well as a resonant sterile neutrino production can take place, depending on temperature and chosen parameters, see \cite{Foot:1995bm,Chu:2006ua}.
This reduces the sterile neutrino contribution to the effective number of relativistic neutrinos $N_{\rm eff}$, see Eq.~\eqref{neff}.
However, the active-sterile flavor conversions take place at later stages and this produces significant distortions in the electron (anti)neutrino spectra, which increase the $^4$He abundance in primordial nucleosynthesis \cite{Saviano:2013ktj}.

Another possibility is to introduce sterile self-interaction processes, the so called {\it secret} interaction model.
In this model the sterile states are coupled to a new $U(1)$ gauge boson \cite{Hannestad:2013ana,Dasgupta:2013zpn} or to a new pseudoscalar \cite{Archidiacono:2014nda,Archidiacono:2015oma,Archidiacono:2016kkh} with a mass much smaller than the $W$ boson.
This new interaction induces a temperature dependent matter potential which suppresses the active-sterile mixing in the early universe and so their relic abundance.
Yet, also in this scenario, significant distortions may be produced in the electron (anti)neutrino spectra, altering the abundance of light element during BBN, see \cite{Mirizzi:2014ama,Saviano:2014esa,Chu:2018gxk}.

\section{Beta decay and neutrino capture}
\label{sec:beta}
The PTOLEMY approach to detect the CNB exploits the neutrino capture processes on $\beta$-unstable nuclei \cite{Weinberg:1962zza,Cocco:2007za,Cocco:2009rh}, like the one with tritium
\begin{equation}
\nu_e + {^{3}\mathrm{H}} \rightarrow {^{3}\mathrm{He}} + e^{-}
 \label{eq:nu-capture-reaction}.
\end{equation}
In fact, tritium has been chosen among other target candidates because of its availability, lifetime, high neutrino capture cross section and low $Q$ value \cite{Cocco:2007za}.
The smoking gun signature of a relic neutrino capture is a peak in the electron spectrum above the $\beta$ decay endpoint.

Because flavor neutrino eigenstates are a composition of mass eigenstates with different masses, while propagating, relic neutrinos quickly decohere into those, in a time scale less than one Hubble time \cite{Weiler:1999ny}.
Therefore, the capture rate of relic neutrinos by  tritium nuclei
\begin{equation}\label{eq:nucapture_events}
\Gamma_{\mathrm{CNB}}
=
\sum^{N_\nu}_{i=1}\Gamma_{i}\,,
\end{equation}
must be computed from the capture rates, $\Gamma_{i}$, of the different neutrino mass eigenstates $\nu_i$:
\begin{equation}
\Gamma_{i}= N_T \,
|U_{ei}|^2 \, \int \frac{d^3 p_\nu}{(2 \pi)^3} \sigma(p_\nu)\, v_\nu f_{\nu_i}(p_\nu)\,,
\end{equation}
where $N_T = M_T / \mH$ represents the number of tritium nuclei in a sample of mass $M_T$ of this element,
$U_{ei}$ are the mixing matrix elements,
$p_\nu$ is the neutrino momentum,
$v_\nu$ is the neutrino velocity as measured at Earth,
$\sigma(p_\nu)$ is the momentum-dependent cross section and
$f_{\nu_i}(p_\nu)$ is the momentum distribution function of the neutrino eigenstate $\nu_i$.
Since the CNB distribution in the phase space is very narrow, the integral reduces to
\begin{equation}\label{eq:nucapture_events_i}
\Gamma_{i}
=N_T\,
|U_{ei}|^2\,
\bar\sigma\, v_\nu
f_{c,i}\,
n_{0}\,,
\end{equation}
where $n_{0}$ is the average neutrino number density on large scales, see Eq.~\eqref{n0},
and $f_{c,i}$ is the clustering factor, that is the local overdensity of these particles due to the gravitational attraction of our galaxy \cite{Ringwald:2004np,deSalas:2017wtt,Zhang:2017ljh}.
The quantity $\bar{\sigma}$ represents the average cross section for neutrino capture,
\begin{equation}\label{eq:crosssection}
\bar\sigma
=
\frac{G_F^2}{2\pi v_\nu}
%|V_{ud}|^2
F(Z,E_e)
\frac{\mHe}{\mH}
E_e p_e
\left(|F|^2+g_A^2 |GT|^2
\right),
\end{equation}
where $\mHe \approx 2808.391\,{\rm MeV}$ and $\mH \approx 2808.921\,{\rm MeV}$ are the \emph{nuclear}\footnote{
The nuclear masses $\mHe$ and $\mH$ are related to the atomic masses
$M_{^3{\rm He}} \approx 2809.413\,\mathrm{MeV}$ and $M_{^3{\rm H}} \approx 2809.432\,\mathrm{MeV}$ \cite{Myers:2015lca}
according to
$\mHe
=
M_{^3{\rm He}}-2m_e+24.58678$~eV
and
$\mH
=
M_{^3{\rm H}}-m_e+13.59811$~eV.
}
masses of the ${^{3}\mathrm{He}}$ and ${^{3}\mathrm{H}}$ nuclei, respectively,
and $E_e$ ($p_e$) is the electron energy (momentum).
The cross section is written in terms of the ``standard''
Fermi ($F$) and Gamow-Teller ($GT$) matrix elements~\footnote{
For the form factors and the axial coupling we use
$|F|^2\simeq0.9987$,
$|GT|^2\simeq2.788$
and
$g_A\simeq1.2695$~\cite{Schiavilla:1998je}.}.
The Fermi function $F(Z,E_e)$ describes the effect of the Coulomb attraction between a proton and the outgoing electron, which enhances the cross section. In order to account for this effect we use the approximation due to
Primakoff and Rosen \cite{Primakoff:1959a},
\begin{equation}\label{eq:fermifunction}
F(Z,E_e)
=
\frac{2\pi \eta}{1-\exp(-2\pi \eta)},
\end{equation}
where $\eta=Z\alpha E_e/p_e$,
$Z=2$ is the atomic number of $^3$He
and $\alpha=1/137.036$~\cite{Tanabashi:2018oca} is the fine structure constant.

Notice the presence of the mixing matrix element $U_{ei}$ in the partial rate $\Gamma_i$. This is due to the fact that only electron neutrinos intervene in the process~\eqref{eq:nu-capture-reaction}, while relic neutrinos are found in their mass eigenstates. In the usual 3 neutrino parameterization \cite{Tanabashi:2018oca}
\begin{equation}\label{eq:mixingmatrix_ei_4}
|U_{ei}|^2
=
(
c_{12}^2 c_{13}^2
%c_{14}^2
,\;
s_{12}^2 c_{13}^2
%c_{14}^2
,\;
s_{13}^2
%c_{14}^2
)\,,
%s_{14}^2
\end{equation}
where $c_{jk}=\cos\theta_{jk}$ and $s_{jk}=\sin\theta_{jk}$,
being $\theta_{jk}$ the corresponding mixing angle.
In our case, we use
the best fit values
$s_{12}^2=0.32$,
$s_{13}^2=2.16\; (2.22)\times10^{-2}$
for normal (inverted) ordering~\cite{deSalas:2017kay} (see also \cite{Capozzi:2018ubv,Esteban:2018azc}).
%and $s_{14}^2=0\; (0.012)$~\cite{Gariazzo:2018mwd}
%when considering three (four) neutrinos.

Because of the finite experimental energy resolution, the main background to the neutrino capture process comes from the most energetic electrons of the $\beta$ decay of tritium, since they can be measured with energies larger than the endpoint.
To estimate the rate of such background, we need to account for the $\beta$ decay spectrum \cite{Masood:2007rc}
\begin{equation}\label{eq:dgamma_beta_de}
\frac{d\Gamma_\beta}{dE_e}
=
\frac{\bar\sigma}{\pi^2}
N_T
\sum^{N_\nu}_{i=1} |U_{ei}|^2
H(E_e, m_{i})\,.
\end{equation}
Defining $y=\Eendz-E_e-m_i$, with $\Eendz$ the energy at the $\beta$ decay endpoint for massless neutrinos,
\begin{eqnarray}
 H(E_e, m_i)
&&=
\frac{1-m_e^2/(E_e \mH)}{(1-2E_e/\mH + m_e^2/\mH^2)^2}
\sqrt{y\left(y+\frac{2m_i\,\mHe}{\mH}\right)} \times \nonumber \\
&& \times \left[y+\frac{m_i}{\mH}(\mHe+m_i)\right]\,.\label{eq:H_function}
\end{eqnarray}
To account for the experimental energy resolution $\Delta$, we
introduce a smearing in the electron spectrum.
This is done using a convolution of both the CNB signal and the $\beta$ decay spectrum
with a Gaussian of full width at half maximum (FWHM) given by $\Delta$.
The smeared neutrino capture event rate $\widetilde\Gamma_{\mathrm{CNB}}$ then reads
\begin{equation}\label{eq:dgamma_nc_de}
\frac{d\widetilde\Gamma_{\mathrm{CNB}}}{dE_e}(E_e)
=
\frac{1}{\sqrt{2\pi}(\Delta/\sqrt{8\ln2})}
\sum^{N_\nu}_{i=1}
\Gamma_i\times
\exp\left\{-\frac{[E_e-(E_{\rm end}+m_i+m_{\rm lightest})]^2}{2(\Delta/\sqrt{8\ln2})^2}\right\}\,,
\end{equation}
where $m_{\rm lightest}$ is the mass of the lightest neutrino and $E_{\rm end}$ is the energy at the $\beta$ decay endpoint, $E_{\rm end} = E_{\rm end,0} - m_{\rm lightest}$.
In the same way, the smeared $\beta$ decays reads
\begin{equation}\label{eq:dgammatilde_beta_de}
\frac{d\widetilde\Gamma_{\beta}}{dE_e}(E_e)
=
\frac{1}{\sqrt{2\pi}(\Delta/\sqrt{8\ln2})}
\int^{+\infty}_{-\infty}
dE'\,
\frac{d\Gamma_\beta}{dE_e}(E')\,
\exp\left[-\frac{(E_e-E')^2}{2(\Delta/\sqrt{8\ln2})^2}\right]\,.
\end{equation}
\begin{figure}
\includegraphics[width=0.5\textwidth]{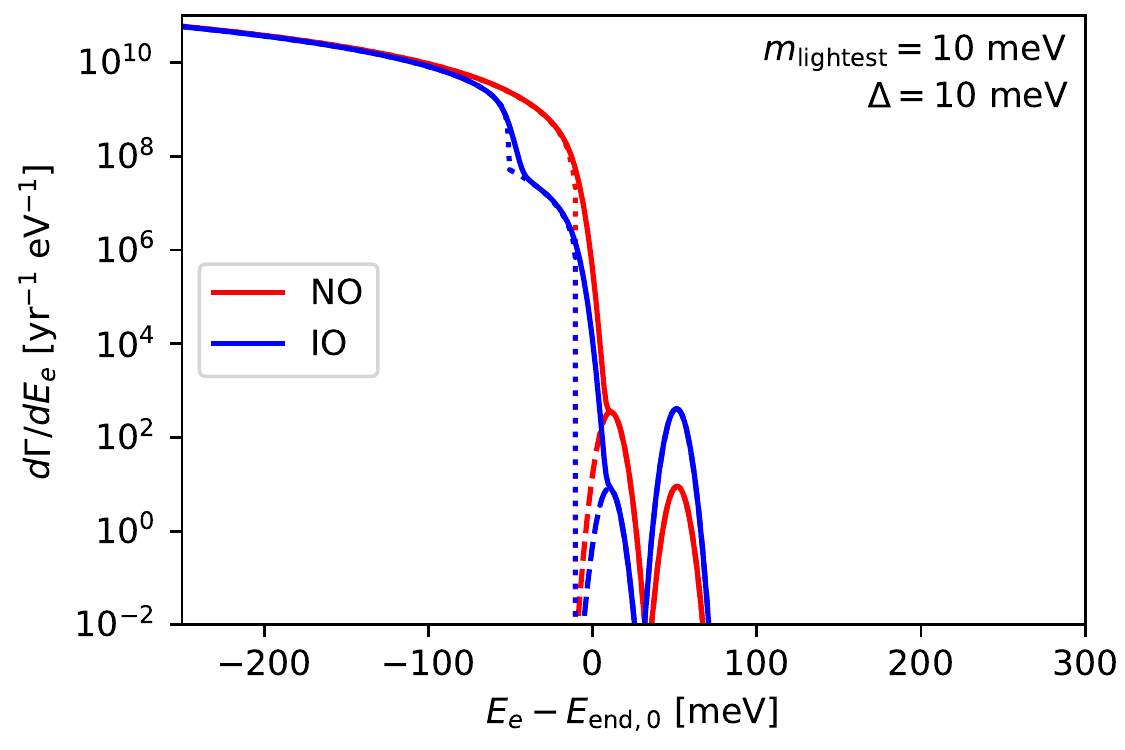}
\includegraphics[width=0.5\textwidth]{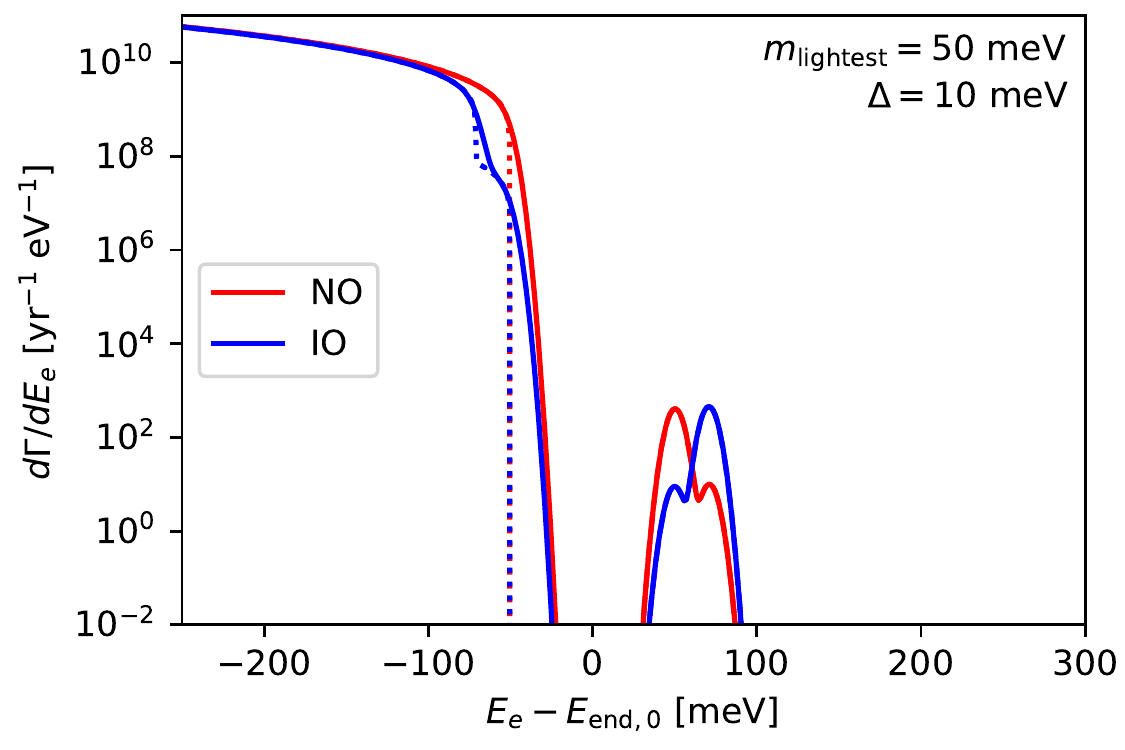}
\includegraphics[width=0.5\textwidth]{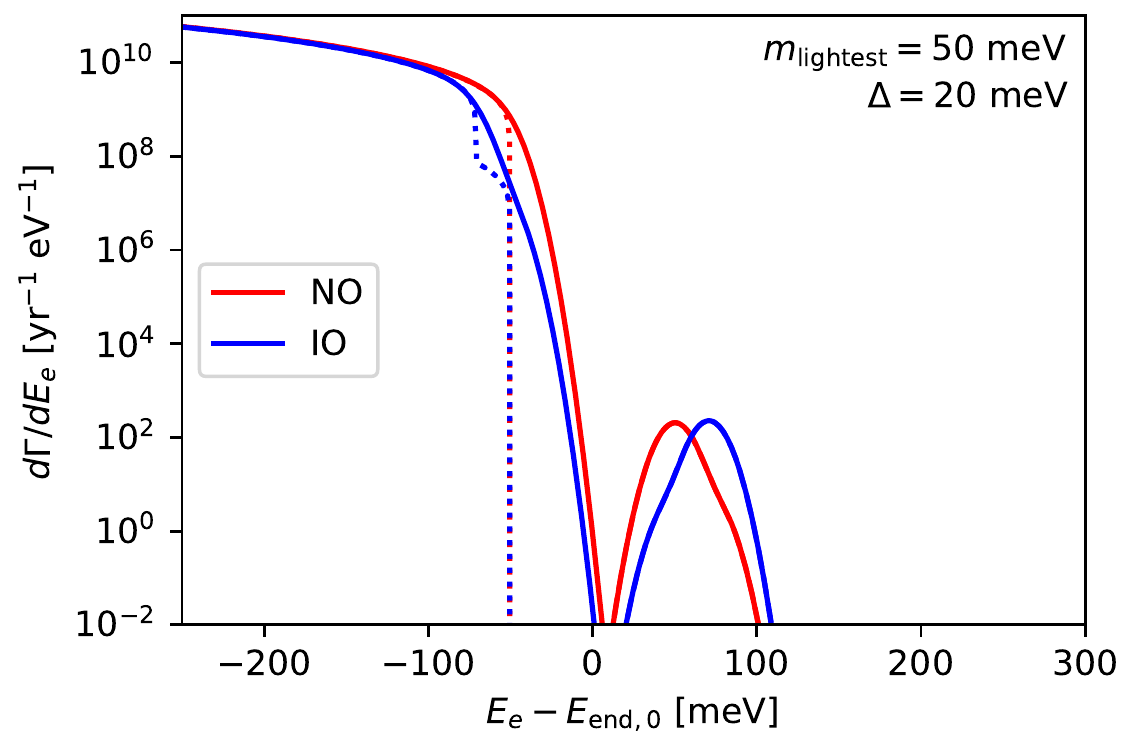}
\includegraphics[width=0.5\textwidth]{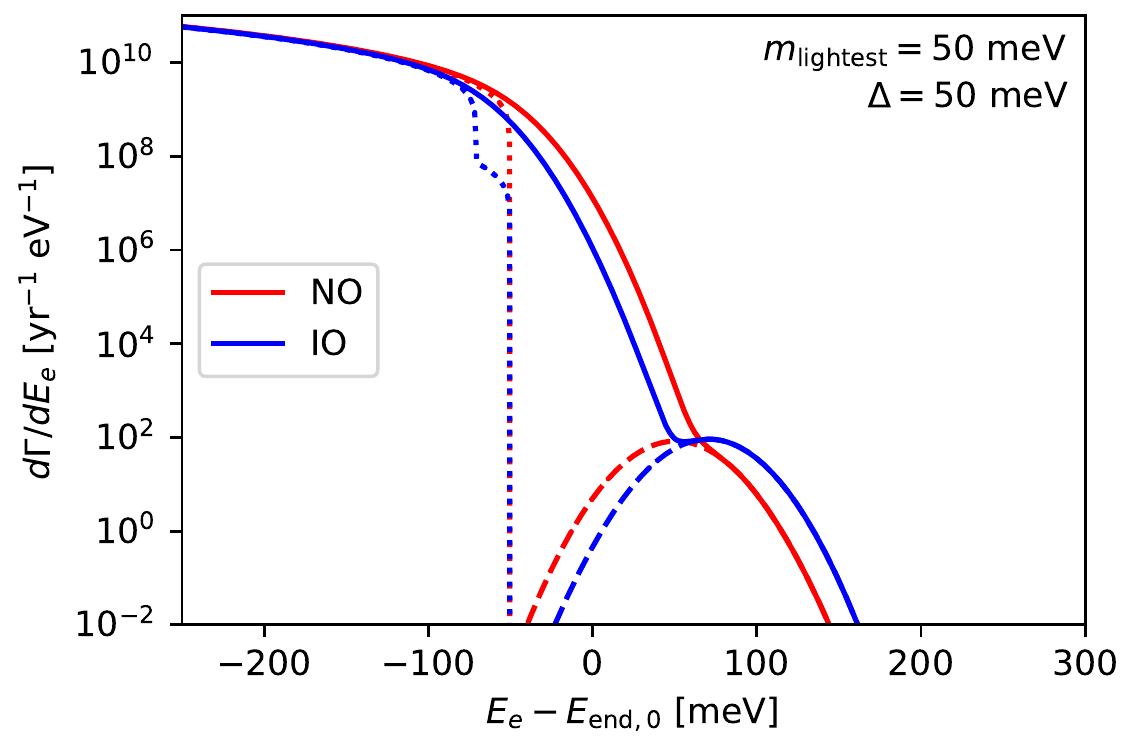}
\caption{Expected event rates versus electron energy $E_e$ in a direct-detection experiment like PTOLEMY (assuming 100~g of tritium source) near the $\beta$ decay endpoint for different lightest neutrino masses and energy resolutions.
Solid lines represent the total event rates convolved with a Gaussian envelope of FWHM equal to the assumed energy resolution, as computed from Eqs.~\eqref{eq:dgamma_nc_de} and \eqref{eq:dgammatilde_beta_de}.
Dashed lines represent the signal event rates as it would be measured by the experiment without the background, while dotted lines show the background ($\beta$ decay) event rates without the convolution, i.e.\ for $\Delta=0$.
Red (blue) lines indicate normal (inverted) ordering.
All lines are obtained considering Dirac neutrinos
and neutrino overdensity according to the semi-analitic expression from \cite{Zhang:2017ljh}.
}
\label{fig:event-rates-beta+CNB-NO-IO}
\end{figure}

The integrated number of signal events is expected to be,
for Dirac neutrinos and ignoring the possible enhancement due to clustering,
around 4 per year \cite{Long:2014zva}.
Depending on the neutrino masses and nature, this number can be enhanced.
As already argued, for Majorana neutrinos the event rate can be a factor two larger,
if neutrinos are massive enough to be all non-relativistic today
or the correct mass ordering is the inverted one \cite{Roulet:2018fyh}.
Concerning the enhancement due to the local relic neutrino density,
it mostly depends on the mass of each mass eigenstate and on the mass ordering.
The event rate may be unaltered if the mass ordering is inverted and the neutrinos are very light,
it could be 10-20\% larger for normal ordering and nearly minimal neutrino masses,
or it may be significantly increased for masses above $\sim$100-150~meV
\cite{deSalas:2017wtt}.

In Fig.~\ref{fig:event-rates-beta+CNB-NO-IO} we show the expected event rates at energies close to the $\beta$ decay endpoint for different neutrino masses and energy resolutions,
and comparing the two possible mass orderings,
considering Dirac neutrinos and taking into account
the neutrino overdensity according to Eq.~\eqref{eq:clust_fac_mi}.
Neutrino capture events can be resolved from the $\beta$ decay background
only for neutrino masses sufficiently larger than $\Delta$, as expected.
In the inverted ordering case (blue) it is interesting to note a kink in the $\beta$ decay spectra due to the larger overlap of $\nu_e$ with the heaviest mass eigenstates.
The feature can be hardly observed in the normal ordering (red) as $\nu_e$ has a much smaller mixing with $\nu_3$.
This is also the reason why the CNB capture peaks have different heights,
proportional to $|U_{ei}|^2$, see Eq.~\eqref{eq:dgamma_nc_de}.
For the smallest energy resolution we have considered, $\Delta = 10\,\mathrm{meV}$,
the contributions due to the different mass eigenstates
($\nu_1$ plus $\nu_2$, and $\nu_3$) can be seen in the upper plots.
In this case, it is also interesting to see that the peak due to the larger mass eigenstate,
$\nu_3$ for normal and $\nu_1$ plus $\nu_2$ for inverted ordering,
has a larger amplitude in the latter case, making the situation more favorable
due to the fact that most of the interesting events have a larger separation from the $\beta$ decay background.

\section{Data analysis method}
\label{sec:analysis}
To estimate the sensitivity of PTOLEMY to the neutrino mass scale we follow and adapt the procedure proposed in the KATRIN Design Report \cite{Angrik:2005ep} and revisited from the Bayesian point of view in \cite{SejersenRiis:2011sj}, see also \cite{Nucciotti:2009wq}.
We consider here in detail the standard active neutrino states, but the analysis can be easily extended to include an extra sterile state with mass in the eV range, see section~\ref{sec:sterile}.
Following the notation adopted in the previous section, we define the number of $\beta$ decay and neutrino capture events within an energy bin centered at $E_i$ as
\begin{eqnarray}
&&N_\beta^i
= T
\int^{E_i+\Delta/2}_{E_i-\Delta/2}
\frac{d\widetilde\Gamma_{\beta}}{dE_e} dE_e\,,
\\
&&N_{\rm CNB}^i
= T
\int^{E_i+\Delta/2}_{E_i-\Delta/2}
\frac{d\widetilde\Gamma_{\rm CNB}}{dE_e}dE_e\, ,
\end{eqnarray}
with $T$ the exposure time.
In our Bayesian simulation we reconstruct the physical parameters given an initial fiducial model.
We will indicate with hats the fiducial parameter values, while the quantities without hats refer to the varying
parameters in the analyses.
For the fiducial models we will select different values for lightest neutrino mass $\hat{m}_{\rm lightest}$,
while the other masses ($\hat{m}_{i}$) and mixing matrix ($\hat{U}$) parameters,
as well as the true endpoint of the $\beta$ spectrum ($\hat{E}_{\rm end}$),
are fixed according to the currently known best fit values
\footnote{When considering the case of sterile states, one should also add a fiducial mass $\hat{m}_{4}$, mixing angle and cosmological number density as suggested by oscillation anomalies and allowed by cosmological data.}.

For the fiducial model, the number of expected events per energy bin is given by:
\begin{equation}
\hat{N}^i
=
N_\beta^i(\hat{E}_{\rm end}, \hat{m}_{i}, \hat{U})
+
N_{\rm CNB}^i(\hat{E}_{\rm end}, \hat{m}_{i}, \hat{U})\,.
\end{equation}
The total number of events that will be measured in a bin is the sum of $\hat{N}^i$ and
a constant background:
\begin{equation}
\hat N_t^i
=
\hat{N}^i
+\hat{N}_b\,.
\end{equation}
Here we will adopt a fiducial PTOLEMY background rate $\hat{\Gamma}_b$,
so that the number of background events becomes $\hat{N}_b= \hat{\Gamma}_b\, T$.
For the main purpose of direct detection of relic neutrinos,
and assuming $\Delta$ of 50 meV,
we require $\hat{\Gamma}_b\simeq10^{-5}$~Hz in the 15~eV region of interest around the endpoint energy,
corresponding to a number of background events of 1 or 2 per each energy bin per year.
This value will be adopted in the following.
Larger background rates may not allow to distinguish
the few signal events that are expected in the full-scale PTOLEMY configuration,
but more detailed studies on the topic are left for future works
where the detector characteristics will be considered in detail.
We then estimate the experimental measurement in each energy bin
using the Asimov dataset, i.e.\ with no statistical fluctuations around the number of events computed
using the fiducial parameter values \cite{Cowan:2010js}:
\begin{equation}
N_{\rm exp}^i(\hat{E}_{\rm end}, \hat{m}_{i}, \hat{U})
=
\hat N_t^i\pm\sqrt{\hat N_t^i}\,,
\end{equation}
assuming a statistical error of $\sqrt{\hat N_t^i}$ in each bin.
Systematic errors will be studied using dedicated Monte Carlo simulations once the detector design
will be more defined.

The simulated measurement is fitted in order to reconstruct the values of the theoretical parameters
that describe the physical model.
We introduce a normalization uncertainty on the number of $\beta$ events ($A_\beta$),
on the endpoint energy ($\Delta E_{\rm end}$) and an unknown constant background ($N_b$).
For these parameters we use linear priors in 
$A_\beta\in[0,2]$, $\ln N_b\in[-1,3]$ and $\Delta E_{\rm end}\in[-1,1]$~eV
and their values will be determined by the fit.
We additionally vary the mass of the lightest neutrino ($\mli\in[0,1]$~eV),
from which we compute the other mass eigenstates according to the mass splittings
measured by current neutrino oscillation experiments, $\dmsq{21}=7.55\e{-5}\text{ eV}^2$
and $\dmsq{31}=2.50\e{-3}\text{ eV}^2$ for normal or $\dmsq{31}=-2.42\e{-3}\text{ eV}^2$ for inverted ordering
\cite{deSalas:2017kay}.

In order to test the perspectives for CNB detection,
we multiply the capture event number by an unknown normalization $A_{\rm CNB}$, whose fiducial value
(i.e.\ the expected value in the standard theoretical scenario) is one, and
for which we consider a linear prior $A_{\rm CNB}\in[0,5]$.
From the fitted value of $A_{\rm CNB}$ one can in principle extract information
on the Dirac/Majorana nature of neutrinos,
on the cross section dependence on NSI
and on the neutrino clustering.
As already mentioned, however, this task will be challenging due to the degeneracy of the various effects
and will not be explored in this work, where we only assess the statistical reach of the PTOLEMY setup.
A direct detection of the CNB (or a measure of the lightest neutrino mass)
at a given C.L.\ can be claimed if the credible interval
for $A_{\rm CNB}$ (or \mli) at that C.L.\ is found to be incompatible with zero.
A more accurate test would require a comparison between the model with free $A_{\rm CNB}$
and the model with $A_{\rm CNB}=0$, for example using the Bayes factor or a maximum likelihood ratio.
We have checked that the results of the two methods are approximately equivalent,
with the model comparison method based on the Savage-Dickey density ratio \cite{Dickey:1971}
giving slightly more pessimistic results.
Since the sensitivity of the PTOLEMY experiment will be more precisely assessed only when
we will know the systematic uncertainties related to the detector,
we do not go in further details here.

For sake of brevity, in the following we will indicate the list of theoretical parameters with
$\bm{\theta} = (A_\beta, N_b, \Delta E_{\rm end}, A_{\rm CNB}, m_i, U)$.
The theoretical number of events in the bin $i$ therefore reads
\begin{eqnarray}
N_{\rm th}^i(\bm{\theta})
&=&
N_b
+ A_\beta\, N_\beta^i(\hat{E}_{end}+\Delta E_{\rm end}, m_i, U) \nonumber
\\
&+&
A_{\rm CNB}\, N_{\rm CNB}^i(\hat{E}_{end}+\Delta E_{\rm end}, m_i, U)
\label{eq:Nth}\,.
\end{eqnarray}
In order to perform the analysis and fit the desired parameters
$\bm{\theta}$,
we use a Gaussian $\chi^2$ function:
\begin{equation}
\chi^2(\bm{\theta})
=
\sum_{i}
\left(
  \frac{N_{\rm exp}^i(\hat{E}_{\rm end}, \hat{m}_{i}, \hat{U})
  -
  N_{\rm th}^i(\bm{\theta})}%
  {\sqrt{N^i_t}}
\right)^2\,,
\end{equation}
which will be converted into a likelihood function $\mathcal{L}$
for the Bayesian analysis according to $\chi^2=-2 \log\mathcal{L}$.
The Gaussian approximation is fully justified for the energy bins for which we have a large number of events,
as expected for the $\beta$ spectrum.
We checked that the presented results for the expected sensitivity on the CNB detection,
which mostly come from bins with a small number of events,
do not change when a Poissonian likelihood is considered instead:
\begin{equation}
\ln\mathcal{L}(\bm{\theta})
=
\sum_{i}
\left(
N_{\rm exp}^i(\hat{E}_{\rm end}, \hat{m}_{i}, \hat{U})\,\ln N_{\rm th}^i(\bm{\theta})
- N_{\rm th}^i(\bm{\theta})
- \ln\Gamma[N_{\rm exp}^i(\hat{E}_{\rm end}, \hat{m}_{i}, \hat{U})+1]
\right)
\,.
\end{equation}

In the following series of simulations, we consider three possibilities for the detector mass.
The full-scale PTOLEMY detector, aiming at the direct detection of the CNB,
requires a tritium mass of at least 100~g, otherwise the signal event rate would be too small to be measurable.
Such an amount of tritium is above the reach of current technology: the first phases of the experiment
will therefore have a less ambitious goal, exploiting a smaller mass of tritium
to test that the available techniques may allow to reach the final target.
In order to demonstrate that the initial phases of the PTOLEMY project will offer interesting opportunities
for studying neutrino properties, such as their mass and possibly the mass hierarchy,
we will present some of the results considering both 1~g and 0.01~g of tritium mass.
We adopt one~year of data taking, an observed energy range between $\hat{E}_{\rm min}=E_0-5$~eV and $\hat{E}_{\rm max}=E_0+10$~eV
and
a constant background rate $\Gamma_b=10^{-5}$~Hz over the whole energy range.

We have verified that increasing $\hat{E}_{\rm max}$ has no impact on the results
if the observed range is sufficient to cover the CNB events,
while some effect may come from a different $\hat{E}_{\rm min}$.
If $\hat{E}_{\rm min}$ is decreased, the precision in measuring the $\beta$ spectrum
(its normalization and the endpoint) allows to slightly improve
the sensitivity on the neutrino parameters, but this comes at the price of a larger number of events, which might be difficult to handle. On the other hand, an $\hat{E}_{\rm min}$ closer to the endpoint allows to reduce the $\beta$ decay event rate
at the expense of slightly worsening the precision on the neutrino mass determination.
The best value for $\hat{E}_{\rm min}$ will be determined once the technical properties of the apparatus
are defined more precisely.

A final comment is about the constant background rate $\hat{\Gamma}_b$.
For an amount of tritium of 100~g, the number of events expected from the $\beta$ decay is much larger
than the background rate and
the determination of the neutrino masses or the detection
of a putative sterile neutrino will be possible even with a much larger $\hat{\Gamma}_b$.
In other words, a much smaller tritium mass might be sufficient to measure the $\beta$ spectrum over the background and achieve sensitivity to neutrino mass scale, as we will discuss later.
Yet, a 100~g target mass and $\hat{\Gamma}_b \lesssim 10^{-5}$~Hz are crucial to allow a detection of the relic neutrinos,
due to the extremely small cross section.

To perform the analysis, we have adapted the generic Markov Chain Monte Carlo (MCMC) sampler
used in \texttt{CosmoMC} \cite{Lewis:2002ah}.
The theoretical parameters that we will try to reconstruct are:
the lightest neutrino mass \mli,
the normalization of the signal spectrum,
the mass ordering,
and, for scenarios with an extra sterile neutrino state,
the squared mass difference $\dmsq{41}\in[0, 20]\text{ eV}^2$ and
the mixing angle $s^2_{14}=|U_{e4}|^2\in[0, 1]$.

\section{Neutrino mass sensitivity and mass ordering}
\label{sec:mass}
\begin{figure}[t]
\centering
\includegraphics[width=0.6\textwidth]{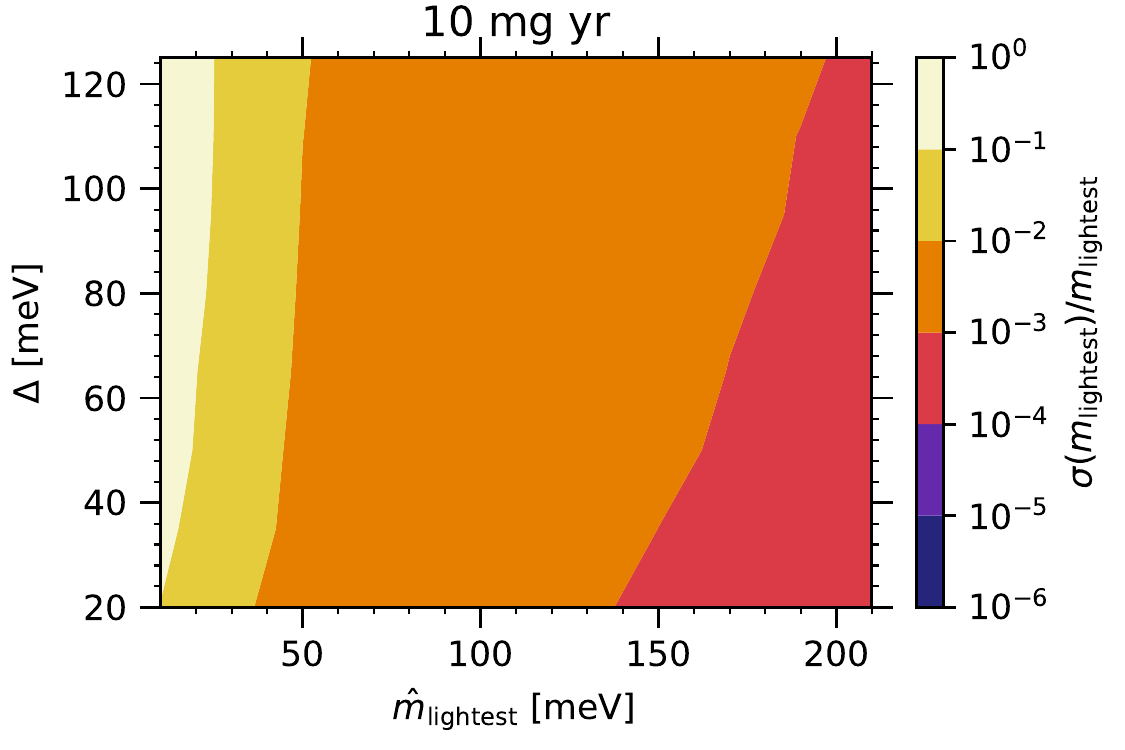}

\includegraphics[width=0.6\textwidth]{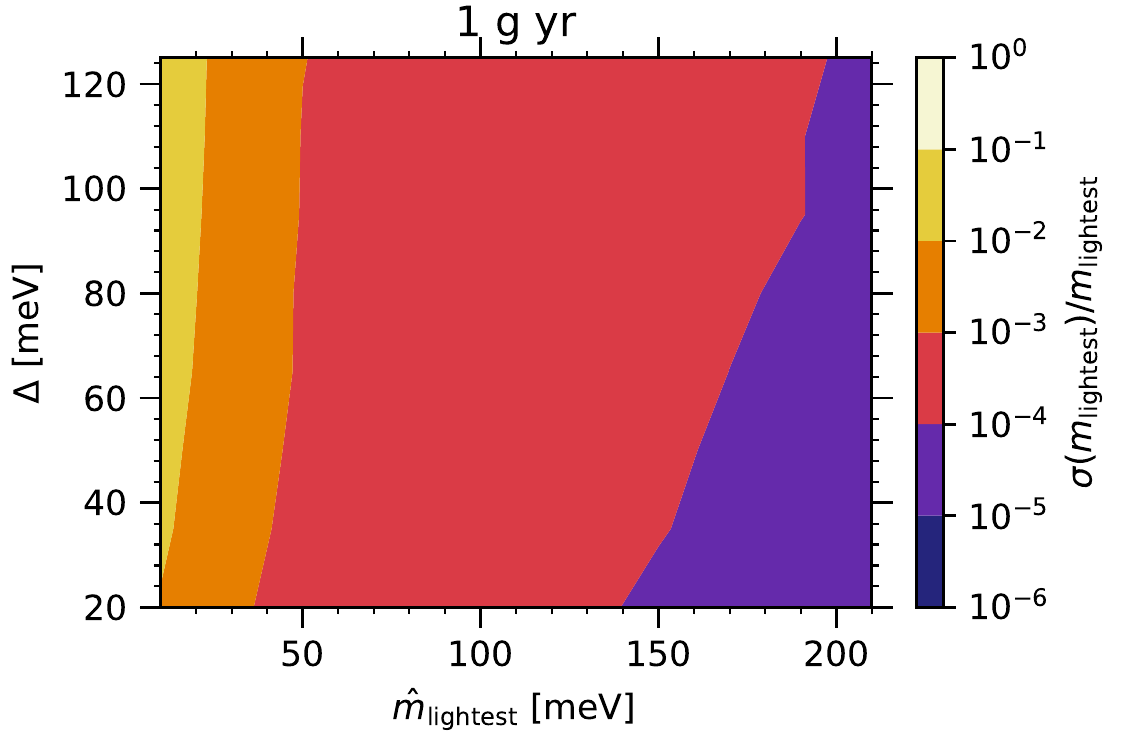}

\includegraphics[width=0.6\textwidth]{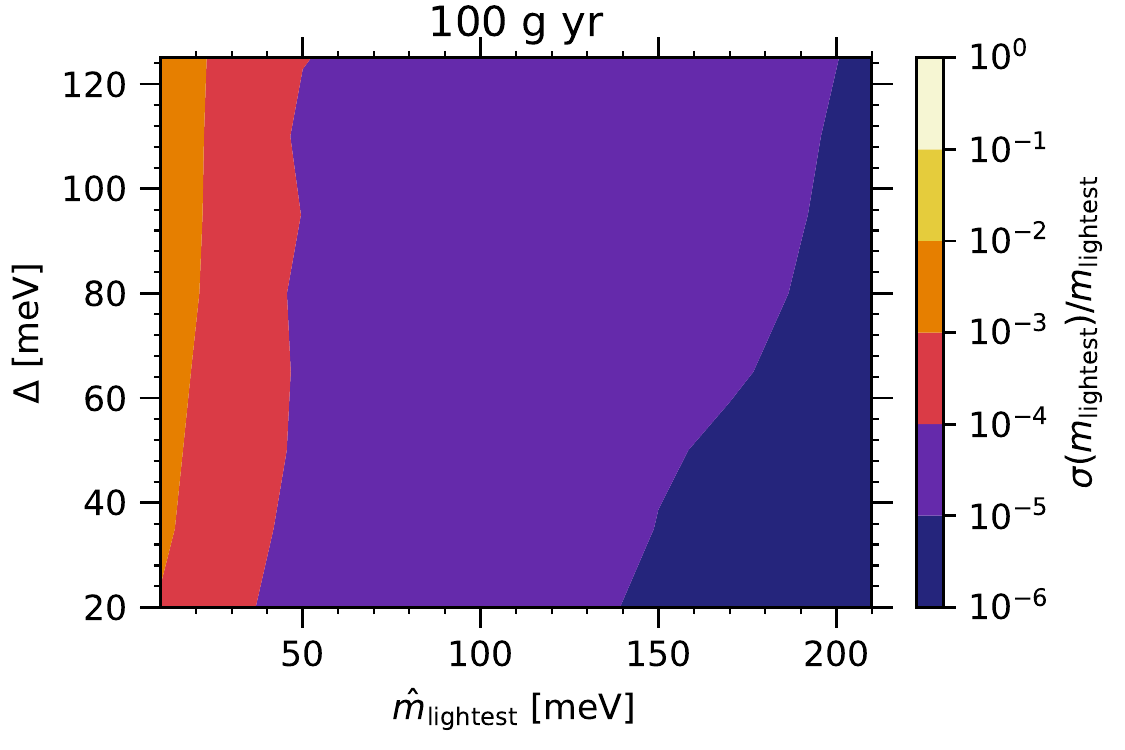}
\caption{Relative error on the reconstructed lightest neutrino mass $m_{\rm lightest}$
as a function of the fiducial lightest neutrino mass $\hat m_{\rm lightest}$ and the energy resolution $\Delta$,
considering 10~mg~yr (top), 1~g~yr or 100~g~yr (bottom) of PTOLEMY data and normal ordering.
The plots for the inverted ordering case are not shown, but are very similar.
}
\label{fig:mass_sensitivity_1y}
\end{figure}

The full-scale PTOLEMY experiment is expected to have an impressive performance in reconstructing the fiducial value
for the lightest neutrino mass, thanks to the large amount of tritium and the correspondingly large statistics.
The $1\sigma$ statistical error obtained from the simulations is of the order of $10^{-3}$~eV or below,
with minimal dependences on the detector configurations described in \cite{Baracchini:2018wwj}, energy range and background rate.
Reasonably, the error slightly depends on the energy resolution of the experiment and on the value of the fiducial lightest neutrino mass, with smaller relative errors for larger masses.
This can be seen in Fig.~\ref{fig:mass_sensitivity_1y},
where we show relative statistical errors obtained
when reconstructing a given fiducial lightest neutrino mass $\hat m_{\rm lightest}$,
for different energy resolutions $\Delta$
and 100~g~yr of PTOLEMY data (lower panel).
As we can see, PTOLEMY may distinguish the neutrino mass
almost independently of the experimental energy resolution,
which has a very small impact on the statistical error on \mli.
The reason is that a larger lightest neutrino mass does not only
induce a shift in the endpoint of the $\beta$ decay spectrum,
but also a change in the normalization of the spectrum at all energies,
which can be measured very well thanks to the very large event rate.

It is worth mentioning that already in the possible initial configurations of the detector, with lower tritium masses,
PTOLEMY may have the ability to measure the neutrino mass.
Considering scenarios with only 10 mg
(upper panel) or 1 g (central panel) of tritium, PTOLEMY has the statistical reach for a determination of the neutrino mass even for $m_{\rm lightest}=10$~meV,
for which a 30\% relative error on the true value of $m_{\rm lightest}$
could be obtained, even with 10~mg of tritium
and in the pessimistic case of $\Delta\sim125$~meV.

Another interesting result that PTOLEMY can obtain is the determination of the neutrino mass ordering.
This is due to the fact that the shape of the $\beta$ spectrum near the endpoint
depends on the single mass eigenstates
and on the mixing matrix elements as described in eq.~\eqref{eq:dgamma_beta_de}
and already shown in Fig.\ \ref{fig:event-rates-beta+CNB-NO-IO}.
Currently, neutrino oscillation data prefer normal ordering (NO, $\dmsq{31}>0$)
over the inverted one (IO, $\dmsq{31}<0$), with a preference of more than 3$\sigma$ \cite{deSalas:2018bym}.
For this reason, we will mostly focus on the NO case.

\begin{figure}[t]
\centering
\includegraphics[width=0.8\textwidth]{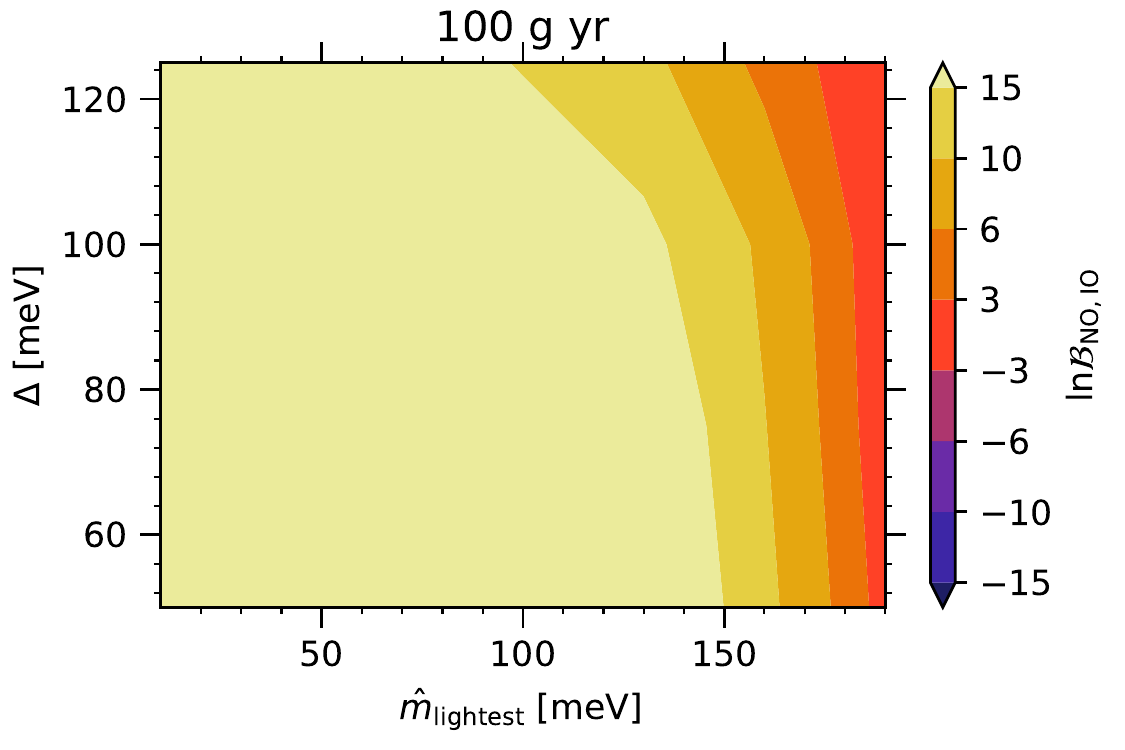}
\caption{Statistical significance for the determination of the neutrino mass ordering,
if the NO is assumed as true,
as a function of
the fiducial lightest neutrino mass
$\hat m_{\rm lightest}$
and the energy resolution $\Delta$,
considering
100~g~yr of PTOLEMY data.
Positive values of $\lnB{\rm{NO,IO}}$ correspond to a preference for NO,
which is statistically decisive ($\gtrsim5\sigma$) if $\lnB{\rm{NO,IO}}\gtrsim15$.
}
\label{fig:order_bayesfac_1y}
\end{figure}

\begin{figure}[t]
\centering
\begin{tabular}{cc}
\includegraphics[width=0.49\textwidth]{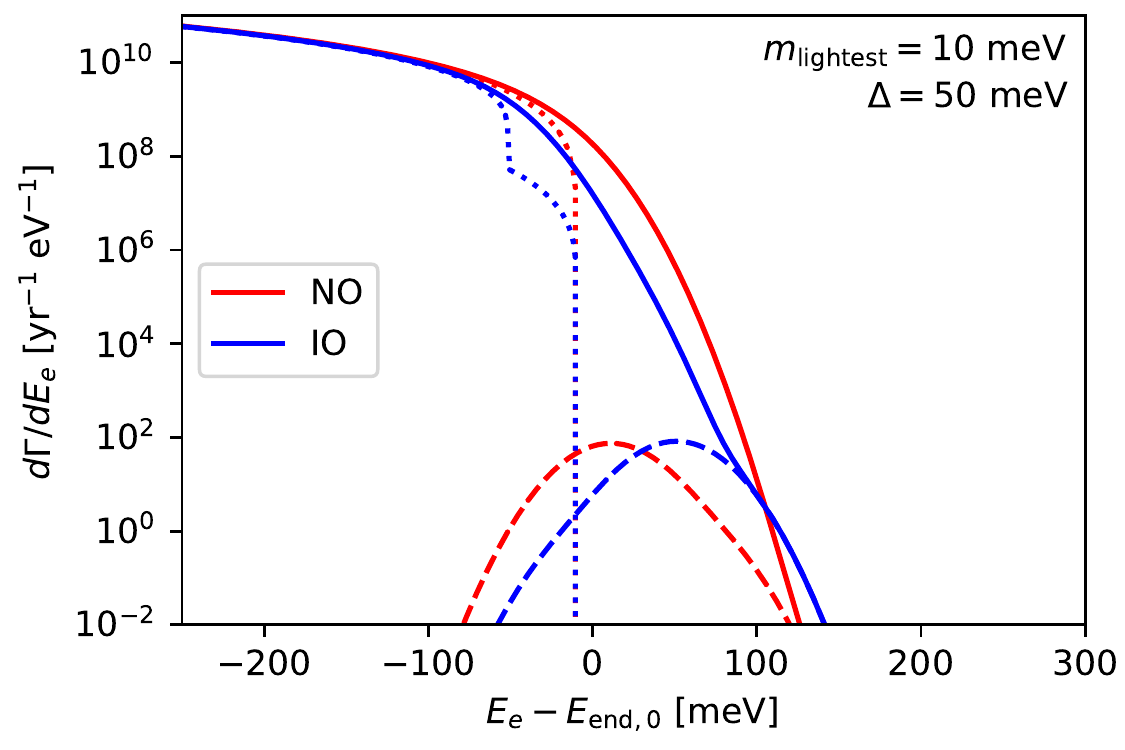}&
\includegraphics[width=0.49\textwidth]{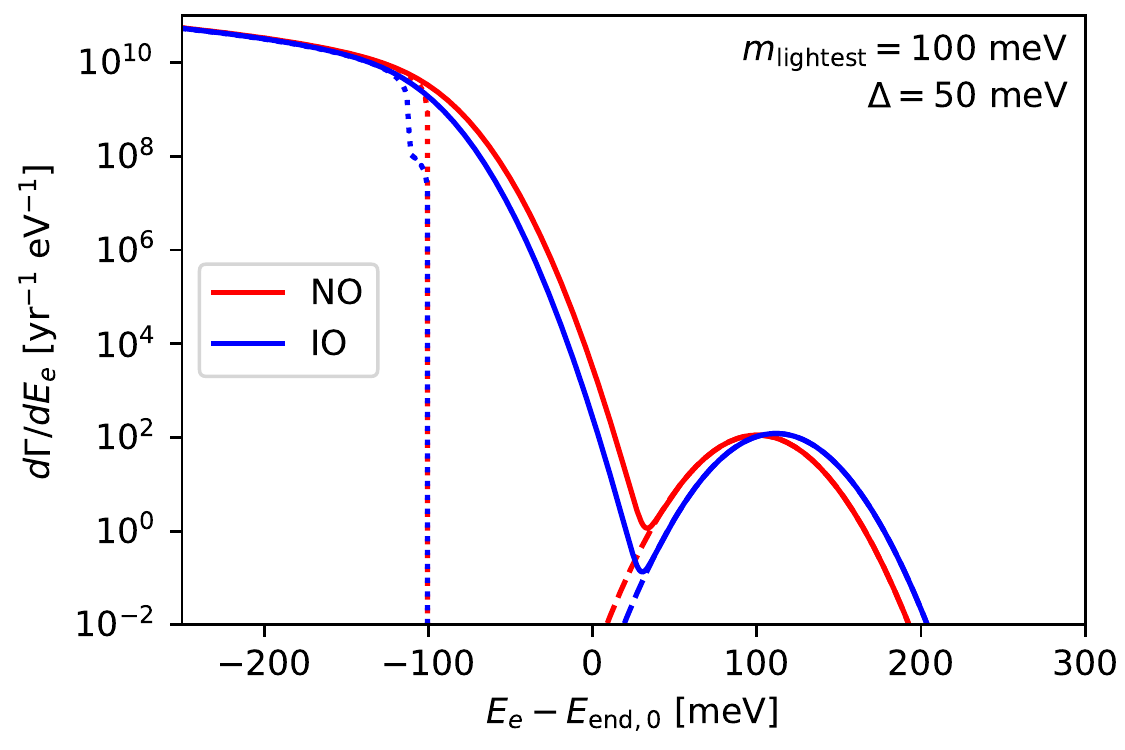}\\
\includegraphics[width=0.49\textwidth]{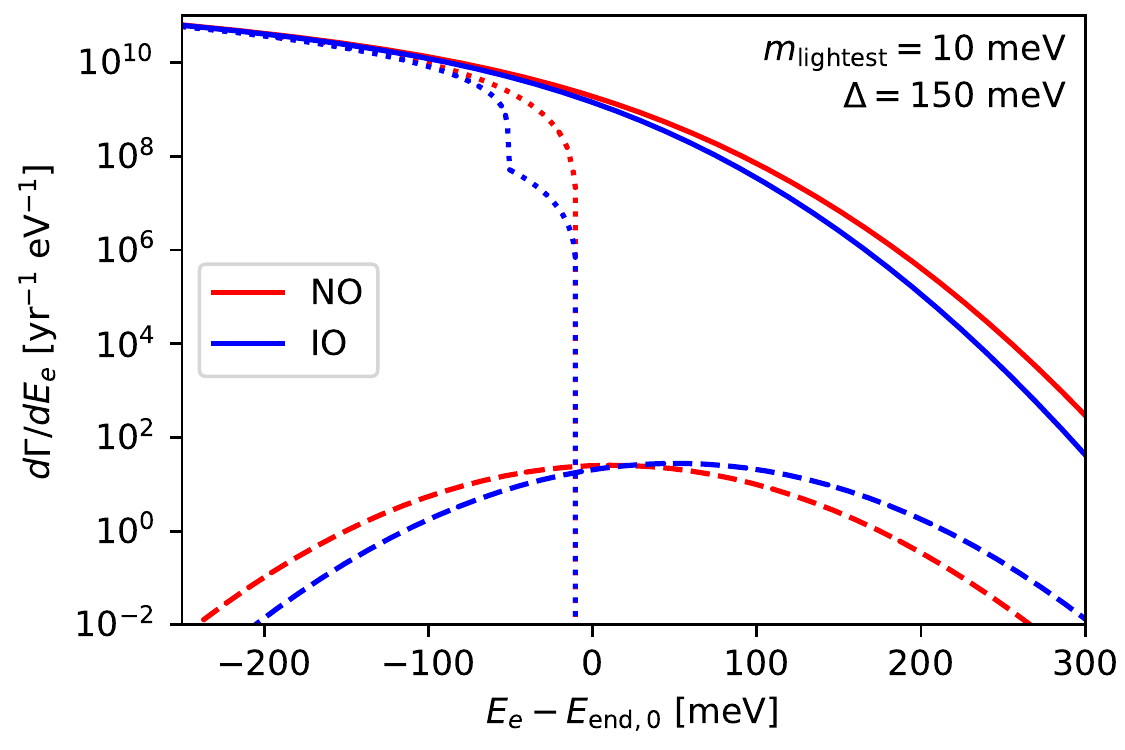}&
\includegraphics[width=0.49\textwidth]{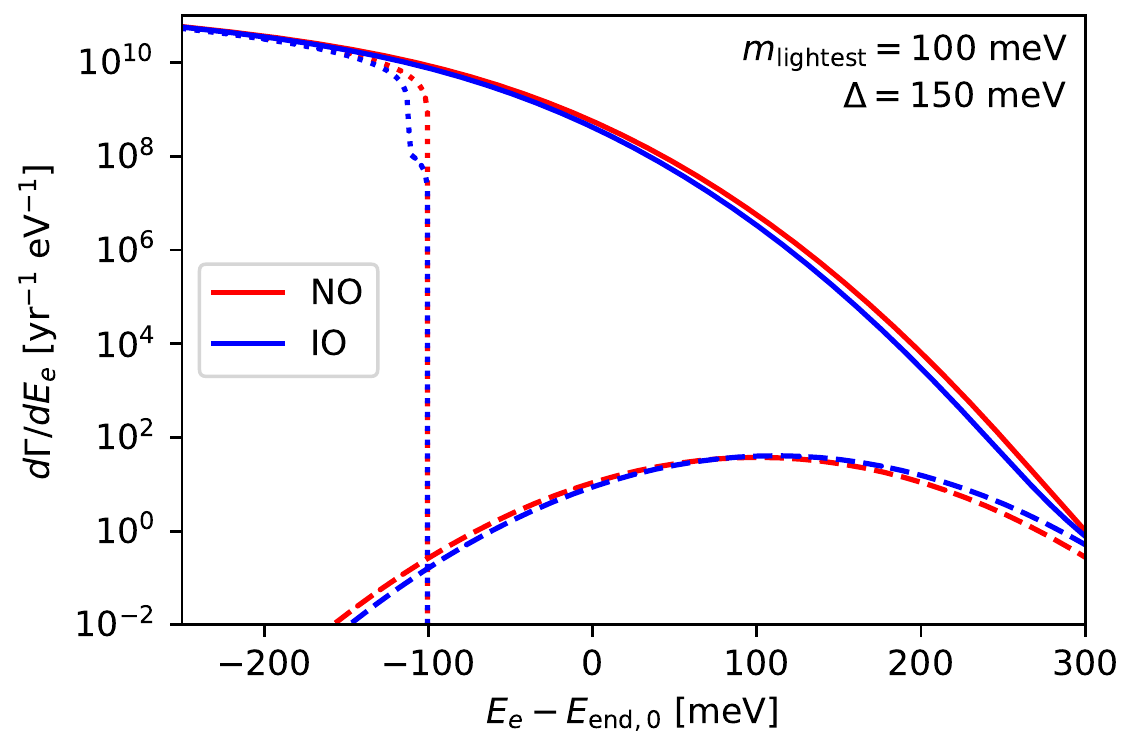}
\end{tabular}
\caption{Comparison of the electron spectra obtained assuming normal (red) or inverted (blue) ordering,
for different values of
the fiducial lightest neutrino mass
$\hat m_{\rm lightest}$ (10 or 100~meV, left to right)
and the energy resolution $\Delta$ (50 or 150~meV, top to bottom),
considering
100~g~yr of PTOLEMY data.
Dashed lines show the signal energy spectrum from neutrino capture, given the considered energy resolution.
Dotted lines indicate the true $\beta$ decay energy spectrum, as it would be measured with a perfect energy determination.
}
\label{fig:spectra_massordering}
\end{figure}

To estimate the capabilities of PTOLEMY in determining the mass ordering,
we assume as fiducial values the best-fit mixing parameters obtained within NO \cite{deSalas:2017kay}.
We then fit the simulated experimental data
using both the NO and IO best-fit mixing parameters and
we compute the Bayesian evidence \bev{}~%
\footnote{For a review on Bayesian model comparison
see e.g.~\cite{Trotta:2008qt},
for its application in determining the neutrino mass ordering see~\cite{Gariazzo:2018pei,deSalas:2018bym}.}.
We will then have two cases:
fiducial NO fitted using NO (\ordfit{NO}{NO} for sake of brevity) and
fiducial NO fitted using IO (\ordfit{NO}{IO}).
The PTOLEMY sensitivity on the mass ordering
is then obtained using the Bayes factor:
\begin{equation}
\label{eq:bayes_factor}
\lnB{ij}=\ln\bev{i}-\ln\bev{j}\,.
\end{equation}
The magnitude of the Bayes factor provides the strength
of the preference for one of the two cases,
while the sign of \lnB{ij} indicates which of the two cases is preferred
(case $i$ if $\lnB{ij}>0$, case $j$ if $\lnB{ij}<0$).
If PTOLEMY is able to distinguish the two orderings,
we expect the fit performed using
the same case as the fiducial choice (\ordfit{NO}{NO})
to be better than the one that assumes
a different ordering with respect to the fiducial one
(\ordfit{NO}{IO}):
\begin{eqnarray}
 \lnB{\rm{NO,IO}}^{\widehat{\rm{NO}}}
 &\equiv&
 \ln\bev{\ordfit{NO}{NO}}-\ln\bev{\ordfit{NO}{IO}}
 >0\,.
\end{eqnarray}
The significance of the preference in favor of NO can be quantified using the absolute value of the Bayes factor.
In terms of its logarithm, the preference for NO is equivalent to a $3\sigma$ probability against IO if $\lnB{\rm{NO,IO}}^{\widehat{NO}}\simeq6$,
to $4\sigma$ if $\lnB{\rm{NO,IO}}^{\widehat{NO}}\simeq10$
or to $\gtrsim5\sigma$ if $\lnB{\rm{NO,IO}}^{\widehat{NO}}\gtrsim15$ \cite{deSalas:2018bym}.
In Fig.~\ref{fig:order_bayesfac_1y} we show $\lnB{\rm{NO,IO}}^{\widehat{NO}}$ as a function of the
fiducial lightest neutrino mass
$\hat m_{\rm lightest}$
and the energy resolution $\Delta$.
As we can see, PTOLEMY will be able to determine the mass ordering (if it is normal)
in the non-degenerate region,
while the distinction will not be feasible ($\lnB{\rm{NO,IO}}$ is inconclusive)
for neutrino masses above $\sim0.18$~eV.
Even if the excellent sensitivity of PTOLEMY for the mass ordering may be unexpected, it has a very simple explanation.
The $\beta$ decay spectrum near the endpoint is significantly different for the normal and inverted ordering cases,
due to the different role of the mixing matrix elements.
The consequence is that when the lightest neutrino has a small mixing with the electron flavor (in the IO case),
the number of events that one can observe close to the endpoint is significantly suppressed.
Here we show only the estimates obtained for 100~g of tritium, but similar conclusions could be obtained when dealing with intermediate-scale experimental configurations.
Considering only statistical errors,
the count rate can change up to two orders of magnitude in the interesting region, as one can see in Fig.\ \ref{fig:spectra_massordering},
where we compare the spectra obtained with different lightest neutrino masses and energy resolutions for normal (red) and inverted (blue) ordering.
As expected, when the mass or the energy resolution are larger the difference between the two spectra diminishes,
but not enough, if the neutrino mass is sufficiently small,
to decrease below the statistical error and
consequently completely lose the sensitivity to the mass ordering.

\section{CNB detection}
\label{sec:cnb}
\begin{figure}[t]
\centering
\includegraphics[width=0.8\textwidth]{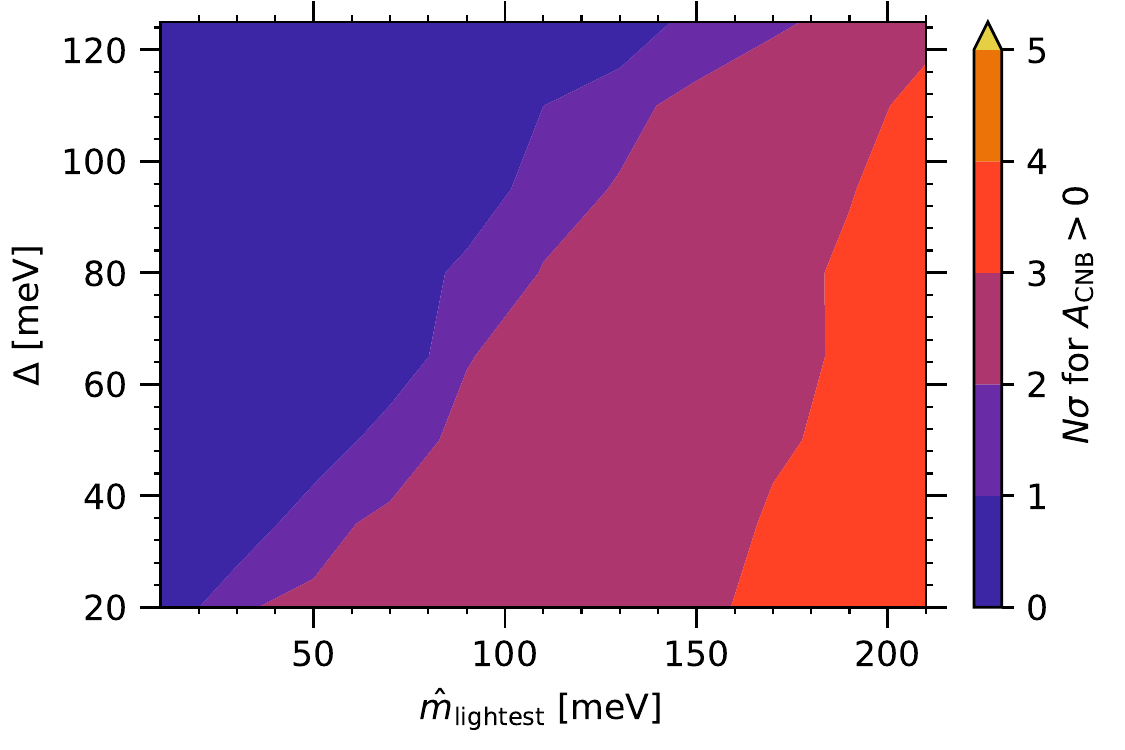}

\includegraphics[width=0.8\textwidth]{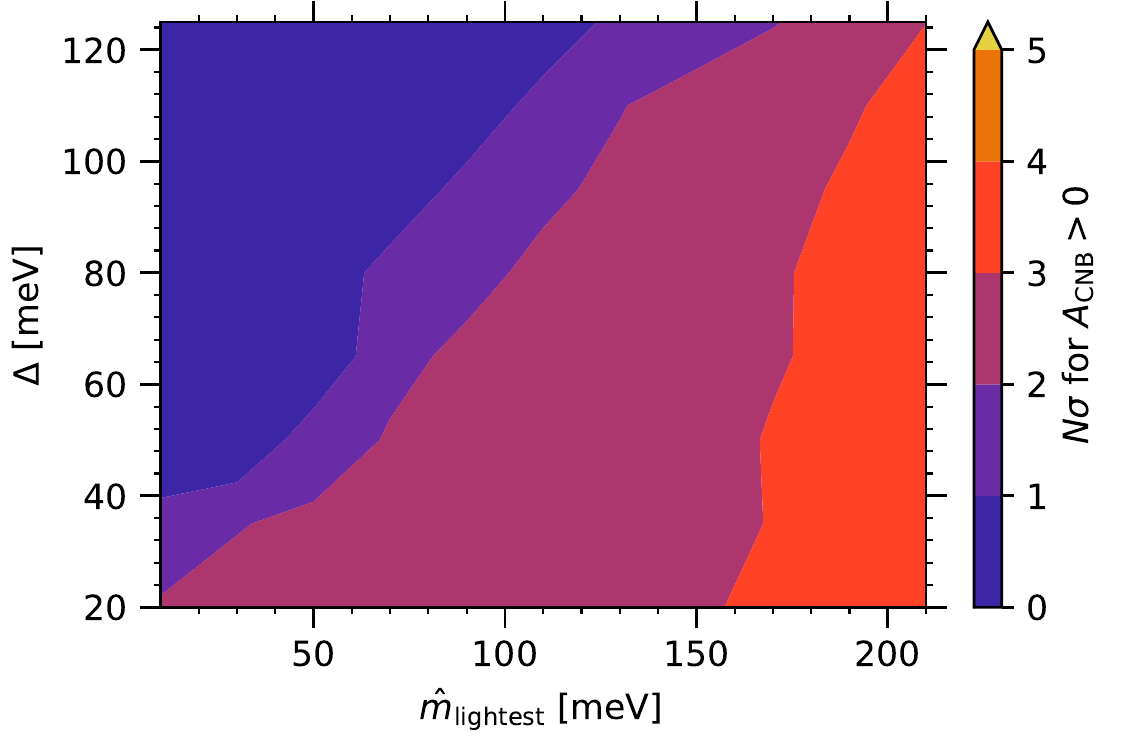}
\caption{Statistical significance for the detection
of the CNB as a function of
the fiducial lightest neutrino mass
$\hat m_{\rm lightest}$
and the energy resolution $\Delta$,
considering
100~g~yr of PTOLEMY data.
The top (bottom) panel represents
normal (inverted) ordering of the neutrino mass eigenstates.
}
\label{fig:Acnb_significance_1y}
\end{figure}

In this Section we investigate the possibility to detect the CNB capture events.
As already mentioned,
we fit the signal from CNB capture
using a free normalization $A_{\rm CNB}$, see eq.~\eqref{eq:Nth},
and we can claim a detection if $A_{\rm CNB}$ can be distinguished
from zero.
Figure~\ref{fig:Acnb_significance_1y} shows the C.L. which can be achieved as a function of the different fiducial lightest neutrino masses
and energy resolutions.
As we can see, it is crucial to achieve a very good energy resolution,
but this may be not enough if the neutrino masses are very small
and the ordering of the mass eigenstates is normal.
While smaller amounts of tritium may be sufficient to study the neutrino mass spectrum,
experimental configurations with less than 100~g of tritium are not suitable for CNB searches,
due to the too low event rate.

The situation does not change significantly if one takes into account the possible enhancement of the event rate
due to the clustering of relic neutrinos in the local dark matter halo, or other effects that could increase
the cross section of the process, such as a Majorana nature of neutrinos or the presence of NSI.
The crucial point, in fact, is that these factors could help to increase the number of observed signal events
only if the energy resolution allows to distinguish them from the $\beta$ decay background,
which has a many orders of magnitude larger rate.

We already noticed that direct detection of relic neutrinos is generally easier
for inverted than for normal mass ordering,
when the neutrino masses are small and the energy resolution is sufficiently good.
This is due to the fact that the primary CNB peaks are shifted at higher electron energies,
because $m_1$ and $m_2$ are larger (see Fig.~\ref{fig:event-rates-beta+CNB-NO-IO}).
As a result, the perspectives of CNB detection
at small neutrino masses are slightly improved in IO with respect to NO,
see the bottom panel of Fig.~\ref{fig:Acnb_significance_1y}.

\section{Sterile neutrinos}
\label{sec:sterile}
In this last section we consider the perspectives for detection of a putative light sterile neutrino
at the eV scale as introduced in Section~\ref{sec:theory}.
It is interesting to scrutinize the information that can be obtained from the distortions
they induce on tritium decay spectrum as well as the effect of their capture.
For the latter case the feasibility of a direct measurement of cosmological sterile states is strongly related to the theoretical model under consideration for the thermalization,
which determines the average number density of the fourth mass eigenstate that appears in Eq.~\eqref{eq:nucapture_events_i},
and to the calculation of the clustering effect of the fourth neutrino, which might be quite large \cite{deSalas:2017wtt}.
Since the thermalization according to standard oscillations is disfavored \cite{Gariazzo:2019gyi},
we will not explore the possibility of achieving a direct detection of relic $\nu_4$ with PTOLEMY in this work.

On the other hand, measuring the $\beta$ spectrum would be extremely useful to constrain the
new squared mass difference \dmsq{41} and mixing angle $s^2_{14}$ through the suppression of the spectrum at energies above $\sim E_0-\sqrt{\dmsq{41}}$,
emerging from eq.~\eqref{eq:dgamma_beta_de}.
Considering the best-fit results from \cite{Gariazzo:2018mwd,Dentler:2018sju} as fiducial values in our analysis,
we find that PTOLEMY would be able to confirm the presence of the sterile neutrino or reject its existence in case of no observation.
Considering a fiducial model with $\dmsq{41}=0$ and $s^2_{14}=0$,
we can get a marginalized $3\sigma$ limit $s^2_{14}\lesssim10^{-4}$ on the relevant mixing angle.
In this case it is very useful to be able to measure a larger fraction of the $\beta$ decay energy spectrum,
since the suppression corresponding to the fourth neutrino starts to be relevant at energies $\sim E_0-\sqrt{\dmsq{41}}$, and what is crucial, thus, is to determine the normalization below and above
this point. This helps to discriminate the contribution of the sterile neutrino from other effects.

\begin{figure}[t]
\centering
\centering
\includegraphics[width=0.6\textwidth]{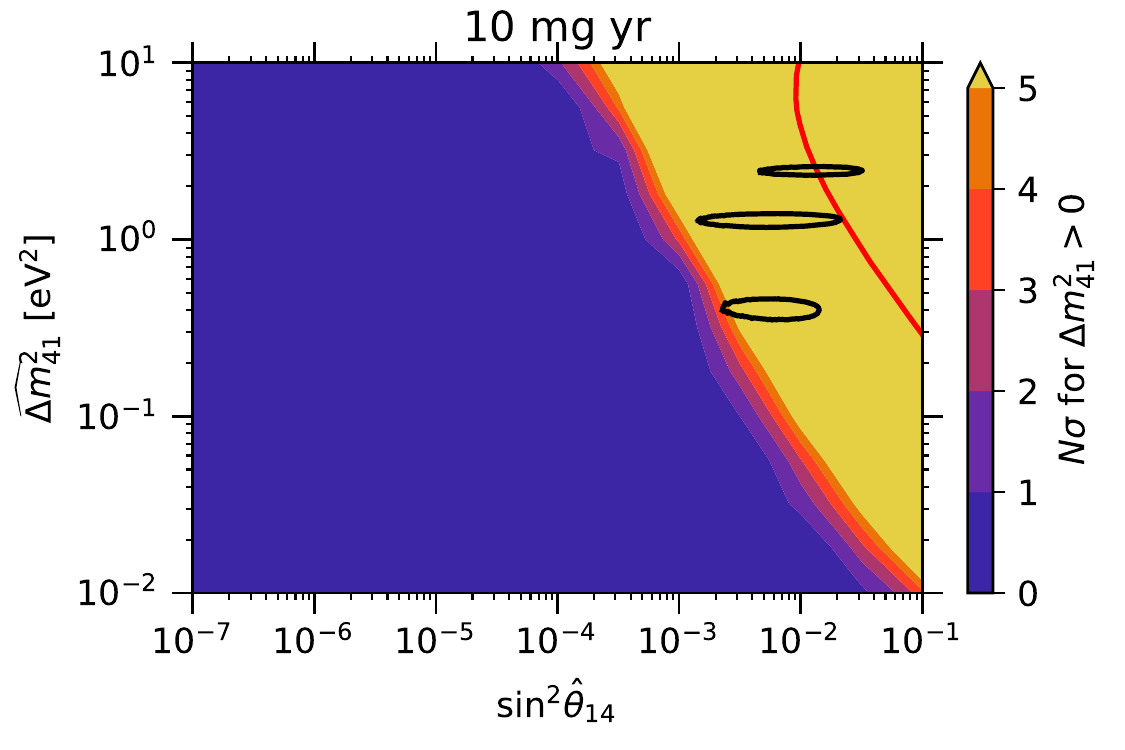}

\includegraphics[width=0.6\textwidth]{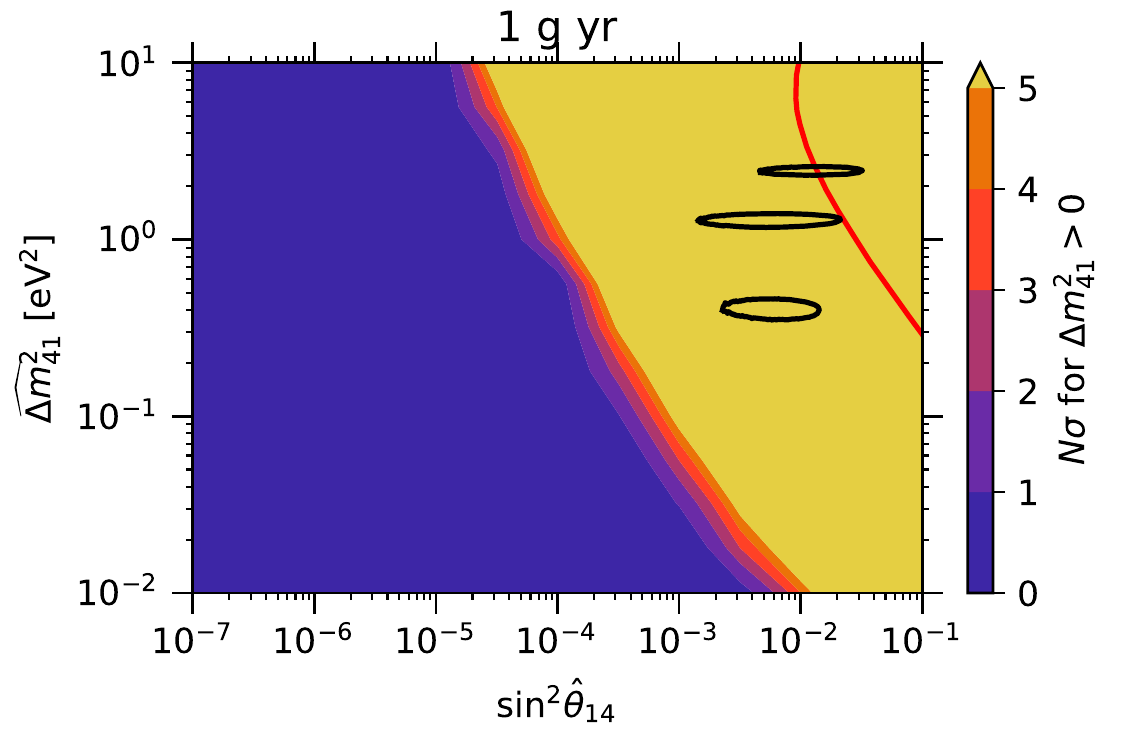}

\includegraphics[width=0.6\textwidth]{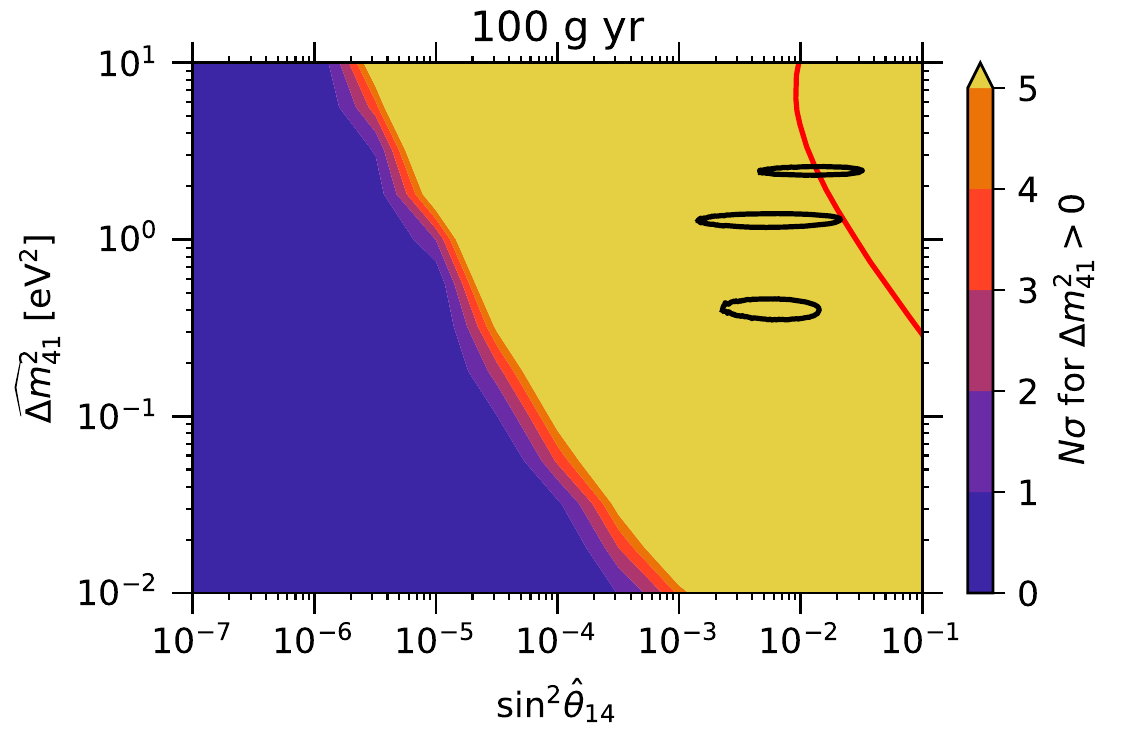}
\caption{
\label{fig:nu4_dm41_significance}
Statistical significance for a detection of \dmsq{41} from measurements of the $\beta$ spectrum,
assuming various fiducial values for the new squared mass difference and mixing angle,
considering
10~mg~yr (top panel), 1~g~yr or 100~g~yr (bottom panel) of PTOLEMY data.
Black contours denote the $3\sigma$ constraints from NEOS and DANSS \cite{Gariazzo:2018mwd},
while the red line shows the 90\% CL sensitivity which is expected for KATRIN \cite{Drexlin:NOW16}.
}
\end{figure}

\begin{figure}[t]
\centering
\centering
\includegraphics[width=0.6\textwidth]{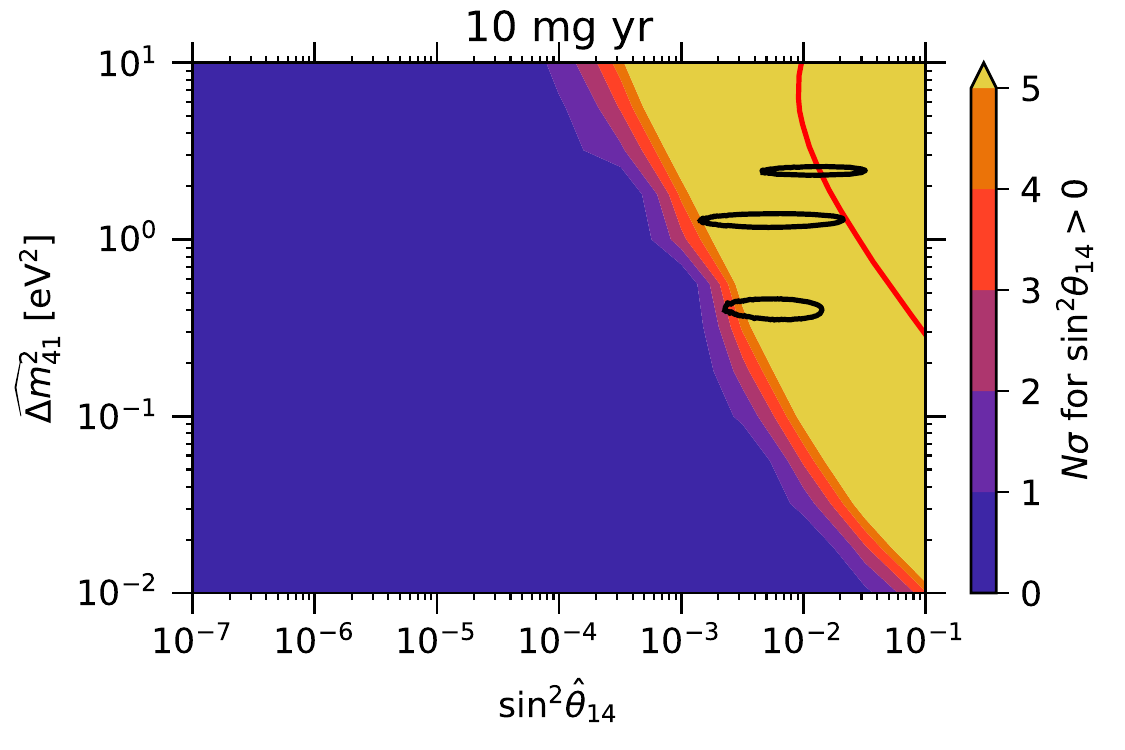}

\includegraphics[width=0.6\textwidth]{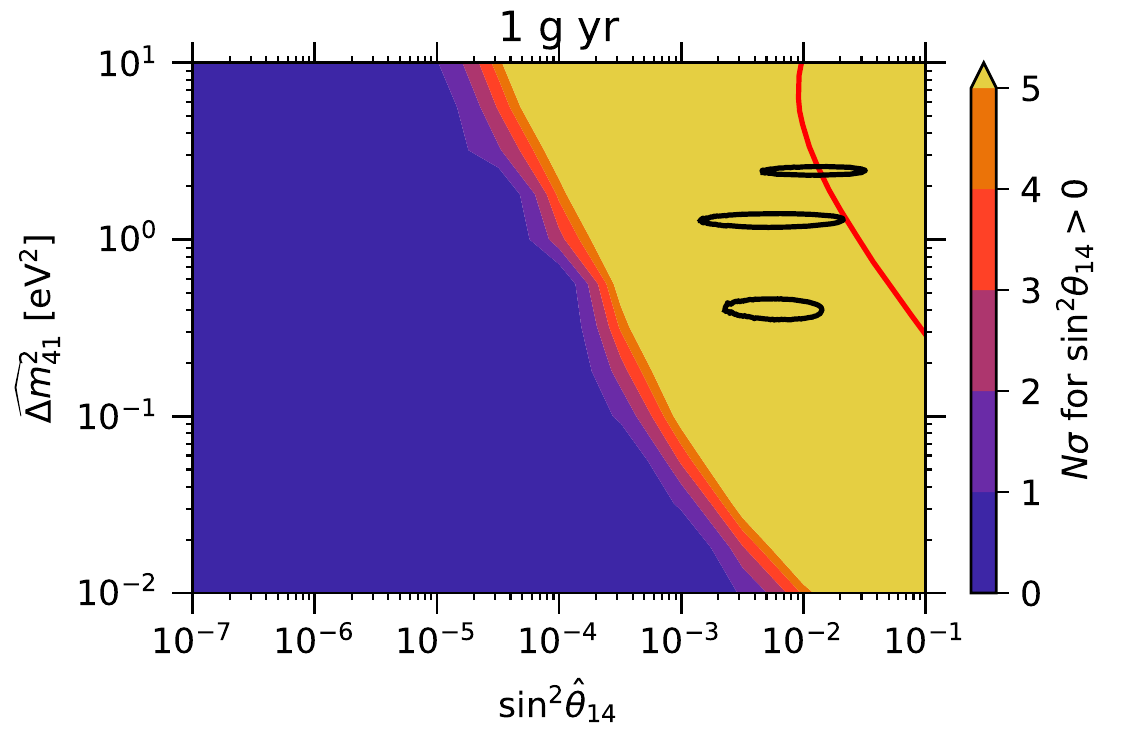}

\includegraphics[width=0.6\textwidth]{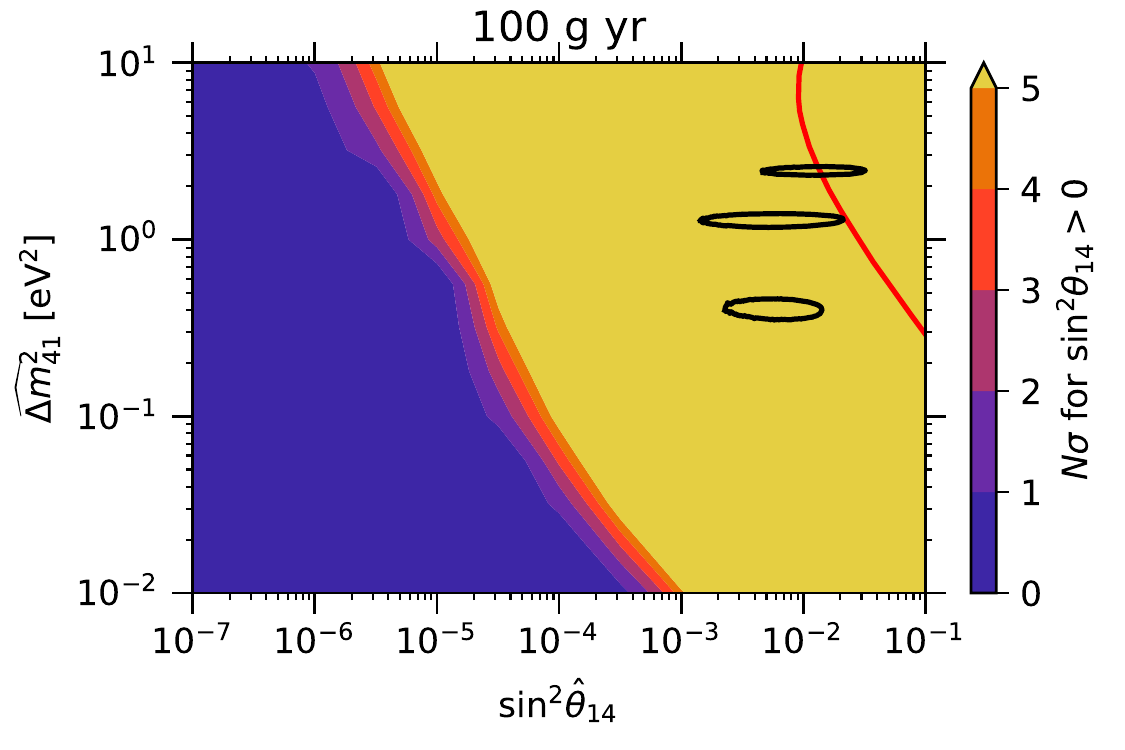}
\caption{
\label{fig:nu4_s14_significance}
The same as Fig.~\ref{fig:nu4_dm41_significance}, but for \ssq{14}.
}
\end{figure}

In Figs.~\ref{fig:nu4_dm41_significance} (for the squared mass difference \dmsq{41})
and \ref{fig:nu4_s14_significance} (for the mixing angle \ssq{14})
we show the prospects for detecting a light fourth neutrino mass eigenstate with a small mixing with the electron neutrino,
considering a wide range of fiducial new squared mass differences and mixing angles.
For very small mixing angles PTOLEMY will hardly distinguish
the effect of a new kink in the $\beta$ spectrum, while
a detection will be clearly possible
given the current preferred values of the best-fit mixing parameters
(see e.g.~\cite{Gariazzo:2018mwd,Gariazzo:2017fdh,Dentler:2017tkw,Dentler:2018sju}),
shown in black in the figure,
even in the case of a small detector with only 10~mg of tritium (top panel).
We also show in red the expected KATRIN sensitivity \cite{Drexlin:NOW16}, which is only marginally covering the present best-fit regions.
The increase in statistics related to the larger tritium mass translates in a better sensitivity also for a light sterile neutrino search.


\section{Conclusions and outlook}
\label{sec:conclusions}
In this paper we have studied the discovery potential of an experiment like PTOLEMY on several aspects of neutrino physics.
The PTOLEMY project has been first conceived as a unique possibility to detect the CNB, a robust prediction of the standard hot big bang model, which has been indirectly proven by several cosmological observables, like BBN and CMB anisotropies.
Yet, since neutrinos are only weakly interacting and because of the very small kinetic energy of the CNB, their direct detection is still an extremely demanding challenge.

In fact, being based on the detection of neutrinos captured by tritium nuclei, a process with no energy threshold, PTOLEMY has a vast physics case, able to detect very low energy fluxes, and to provide constraints on neutrino properties, those of standard ones as well as of exotic sterile species.
Furthermore, it was also realized that the PTOLEMY setup (but without tritium) could be a way to observe dark matter particles with masses in the MeV mass range, keeping track of their arrival directions \cite{Hochberg:2016ntt, Cavoto:2017otc},
while the complete detector setup (with tritium) can also serve as a ``laboratory'' to test ``ab-initio'' theoretical predictions of few-nucleon systems, and
therefore ultimately of the models for nuclear interactions and weak currents \cite{Marcucci:2018baw}.

Our analysis has been focused on three issues.
What is the PTOLEMY sensitivity to the neutrino mass scale?
What is its discovery potential of CNB?
What are the prospects to constrain (or exclude) extra sterile neutrino states which are mixed with standard ones, and have a mass in the eV range?
Exploiting a Bayesian approach, we have analyzed the role of the two main parameters which affect the sensitivity of a PTOLEMY-like experiment, namely the emitted electron energy resolution and the tritium sample mass, forecasting an answer to each of the questions above.

The absolute neutrino mass scale is still an unknown parameter, though cosmological observations and present results from tritium decay constrain the sum of the neutrino masses to be less than $0.12-0.6$ \cite{Ade:2015xua,Lattanzi:2017ubx,Gariazzo:2018meg,Aghanim:2018eyx} or 6~eV \cite{Tanabashi:2018oca}, respectively.
The KATRIN experiment is expected to provide a limit which enters into the region for the neutrino mass that is still marginally allowed by cosmology.
In view of the larger tritium mass which would be used, this bound could be further improved by PTOLEMY, at the level of tens of meV or less with a 1~g tritium mass, {\it independently of the electron energy resolution}, provided \mli\ is in the range 10-150 meV.
Moreover, we have also found that there are interesting opportunities to check the neutrino mass hierarchy, from a different and independent
approach than oscillation experiments.
This can be achieved not only with the largest tritium mass considered in this paper, 100~g, but also with intermediate smaller amounts of order of grams or fractions of grams.

Of course the most ambitious goal of PTOLEMY is the detection of the CNB.
We have performed an analysis of the detection sensitivity as a function of the tritium mass and the energy resolution.
A tritium mass of the order of 100~g, unless the neutrino local clustering is much larger than what is obtained by simulations, is required.
We report the discovery potential as function of the lightest neutrino mass and energy resolution, as shown in Fig.~\ref{fig:Acnb_significance_1y}.
As expected, these two parameters are almost linearly correlated: a smaller electron energy resolution implies a better constraint on smaller neutrino mass scales, and, for example, with $\Delta \sim 100$~meV, it would be possible to detect the CNB at $\sim2\sigma$ if $\mli\gtrsim120$~meV.
Despite the demanding technological issues in dealing with a large tritium mass, as well as in achieving high energy resolutions (which, however, are not so far from present values), we note that there are no other feasible approaches to \emph{directly} unveil the CNB.
The only other plausible one, the Stodolsky effect \cite{Stodolsky:1974aq}, is much more challenging to be detected, if not simply impossible, if cosmological neutrinos have zero (or exceedingly small) chemical potential.

We have also considered the case of an extra sterile neutrino state with a mass of order eV and quite large mixing with active neutrinos.
This sterile particle may help in solving the anomalies found in oscillation experiments, though the picture is still puzzling, since appearance and disappearance observations in short baseline experiment are only marginally in agreement once the 3+1 neutrino scenario is assumed \cite{Dentler:2018sju}.
Cosmological data strongly disfavor a relic density of these sterile neutrinos of the order of the standard ones.
Since their contribution to the neutrino capture rate is also suppressed by their mixing angle $\sin^2\theta_{14}\sim 0.012$, it seems difficult that they could be measured by their capture on tritium nuclei.
It is much more promising that the features they would induce on the standard $\beta$ decay spectrum could give more stringent constraints on their mass and mixing angles.
In particular, we found that a sterile neutrino ``detection'' could be achieved by PTOLEMY, given the current preferred values of the best-fit mixing parameters,
{\it even in the case of a small detector with only 10~mg of tritium}.
In case of no observation, strong limits on the mixing angles will be derived.

In the future we plan to address a similar analysis for the case of a sterile neutrino with mass in the keV range, which has been considered as a warm dark matter candidate.
If we assume that the whole dark matter is made by such particles, their local energy density would be of the order of GeV cm$^{-3}$, corresponding to a rather large local number density of $10^5$~cm$^{-3}$, while their average density on cosmological scales is five orders of magnitude smaller.
This means that the sterile states cannot have been produced in equilibrium in the early universe.
This, together with astrophysical constraints, bounds the sterile-active mixing angle to be very small, $\ssq{i4} \leq 10^{-8}$ \cite{Adhikari:2016bei}.
We thus expect the sterile neutrino capture signal to be too small to be detected by PTOLEMY, while the analysis of the much larger event number expected in the $\beta$ decay spectrum might provide bounds on both $m_4$ and mixing angles.
For this analysis, however, a larger fraction of the $\beta$ spectrum needs to be measured, as an observation around the endpoint is not sufficient to single out the kink corresponding to a mass in the keV range
and the sterile neutrino presence would be degenerate with the normalization of the $\beta$ spectrum.

Another interesting issue, which deserves further studies, is to use PTOLEMY-like
experiments as a way to check the models for nuclear interactions and currents.
The three-nucleon bound systems
involved in the PTOLEMY experiment have been the object of intense theoretical
studies since many years. They represent an ideal ``laboratory'' to
test our understanding of how nucleons interact among themselves,
as well as with external electroweak probes. In order to do so,
the three-body Schr\"odinger equation has to be solved exactly.
Nowadays, a variety of methods exists to this aim: among the most accurate ones we can list the so-called
Hyperspherical Harmonics (HH) method (see Ref.~\cite{Kievsky:2008es} and
references therein). Historically, the models for the nuclear interaction and currents
have been derived within two different frameworks:
a purely phenomenological one, and, more recently, the so-called
chiral effective field theory
approach ($\chi$EFT) (see Ref.~\cite{Machleidt:2016rvv} and references therein).
The theoretical results for the cross section
obtained within this ``ab-initio'' approach for the tritium decay and neutrino capture process will represent predictions
which the PTOLEMY experiment results can check and validate for the unpolarized and, more interestingly, for the polarized rate.

\acknowledgments
Work supported
by the Italian grant 2017W4HA7S ``NAT-NET: Neutrino and Astroparticle Theory Network'' (PRIN 2017)
funded by the Italian Ministero dell'Istruzione, dell'Universit\`a e della Ricerca (MIUR);
by the Spanish grants
ENE2016-76755-R, SEV-2014-0398 and FPA2017-85216-P (AEI/FEDER, UE),
PROMETEO/2018/165 (Generalitat Valenciana),
Mar{\'i}a de Maeztu Unit of Excellence CIEMAT - Particle Physics (MDM-2015-0509)
and the Red Consolider MultiDark FPA2017-90566-REDC;
by the Vetenskapsr{\aa}det (Swedish Research Council) through contract No.\ 638-2013-8993;
by the Simons Foundation, USA (\#377485) and John Templeton Foundation, USA (\#58851);
and by the European Union's Horizon 2020 research and innovation program
under the Marie Sk\l{}odowska-Curie individual Grant Agreement No.\ 796941.

\bibliography{main}

\end{document}